\documentclass[prd,aps,superscriptaddress,twocolumn,nofootinbib,showpacs,preprintnumbers,floatfix]{revtex4}

\pdfoutput=1

\usepackage{graphicx}
\usepackage{dcolumn}
\usepackage{amsfonts,amsmath,amssymb,amsthm}
\usepackage{euscript,bbm}
\usepackage{ifthen}
\usepackage{psfrag}

\usepackage{subdepth}

\usepackage{color}

\usepackage{comment}

\usepackage{bm}


\usepackage{slashed}

\declareslashed{}{/}{0}{-0.05}{E}
\declareslashed{}{/}{0}{-0.05}{P}
\declareslashed{}{/}{0}{-0.10}{\Widehat{P}}

\usepackage{float}
\usepackage{fancybox}
\usepackage{fancyvrb}
\usepackage{ragged2e}
\cornersize*{15pt}
\newfloat{card}{htb}{lop}
\floatname{card}{Card}

\makeatletter
\renewcommand\floatc@plain[2]{\setbox\@tempboxa\hbox{{\footnotesize {\bf {\@fs@cfont #1:}} #2}}%
\ifdim\wd\@tempboxa>\hsize {\begin{minipage}[t]{0.9\linewidth}\justifying \small {\raggedleft {\@fs@cfont #1:} #2}\end{minipage}}\par
\else\hbox to\hsize{\hfil\box\@tempboxa\hfil}\fi}
\@namedef{thecard}{\Alph{card}}
\makeatother

\def\met{{\slashed{E}} {}_{\rm T}}

\def\fsu5{\mbox{$\cal{F}$-$SU(5)$}}
\def\bfsu5{$\boldsymbol{\mathcal{F}}$-$\boldsymbol{SU(5)}$}

\newcommand{\be}{\begin{equation}}
\newcommand{\ee}{\end{equation}}
\newcommand{\bea}{\begin{eqnarray}}
\newcommand{\eea}{\end{eqnarray}}
\newcommand{\nn}{\nonumber}

\newcommand{\gsim}{\lower.7ex\hbox{$\;\stackrel{\textstyle>}{\sim}\;$}}
\newcommand{\lsim}{\lower.7ex\hbox{$\;\stackrel{\textstyle<}{\sim}\;$}}

\newcommand{\ttbar}{t \bar{t}}

\newcommand{\pT} {{P_{\rm T}}}

\newcommand{\PlotPairWide}[4]{
\begin{figure*}[htp]
\centering
\hspace{-5pt}
\includegraphics[width=3.4in]{#1}
\hspace{+10pt}
\includegraphics[width=3.4in]{#2}
\begin{minipage}[l]{0.8\linewidth}\vspace{-10pt} \justifying \small {\raggedleft \caption{#4} \label{#3}}\end{minipage}\vspace{0pt}
\end{figure*}}

\newcommand{\PlotSingle}[4]{
\begin{figure}[htp]
\centering
\includegraphics[width=#2]{#1}
\begin{minipage}[l]{0.8\linewidth}\vspace{-0pt} \justifying \small {\raggedleft \caption{#4} \label{#3}}\end{minipage}\vspace{0pt}
\end{figure}}

\newcommand{\PlotPairStack}[4]{
\begin{figure}[htp]
\centering
\hspace{-5pt}
\includegraphics[width=3.4in]{#1}\\
\vspace{-5pt}
\hspace{-5pt}
\includegraphics[width=3.4in]{#2}
\vspace{-15pt}
\caption{#4}
\label{#3}
\end{figure}}

\begin{document}

\title{Probing the Goldstone equivalence theorem in Heavy Weak Doublet Decays}

\author{Bhaskar Dutta}
\affiliation{George P. and Cynthia W. Mitchell Institute for Fundamental Physics and Astronomy, Texas A$\&$M University, College Station, TX 77843, USA}
\author{Yu Gao}
\affiliation{George P. and Cynthia W. Mitchell Institute for Fundamental Physics and Astronomy, Texas A$\&$M University, College Station, TX 77843, USA}
\author{David Sanford}
\affiliation{Walter Burke Institute for Theoretical Physics, California Institute of Technology, Pasadena, CA 91125, USA}
\author{Joel W. Walker}
\affiliation{Department of Physics, Sam Houston State University, Huntsville, TX 77341, USA}

\begin{abstract}
This paper investigates the decays from heavy higgsino-like weak-doublets into $Z, h$ bosons and missing
particles. When pair-produced at the LHC, the subsequent $Z,h\rightarrow \ell\ell, b\bar{b}$
decays in the doublet decay cascade can yield $4 \ell,2\ell 2b$ and $4b$ + $\met+j$(s) final states.
Mutual observation of any two of these channels 
would provide information on the the associated doublets' decay branching fractions into a $Z$ or $h$,
thereby probing the Goldstone equivalence relation, shedding additional
light on the Higgs sector of beyond the Standard Model theories,
and facilitating the discrimination of various contending models, in turn.
We compare the $Z/h$ decay ratio expected in the Minimal Supersymmetric model, the Next-to
Minimal Supersymmetric model and a minimal singlet-doublet dark matter model.
Additionally, we conduct a full Monte Carlo analysis of the prospects for detecting
the targeted final states during 14 TeV running of the LHC
in the context of a representative NMSSM benchmark model.
\end{abstract}
\preprint{MI-TH-1530, CALT-2015-039}
\maketitle


\section{Introduction\label{sct:intro}}

The Higgs mechanism plays a central role in the electroweak symmetry breaking and many beyond the Standard
Model (BSM) frameworks have been proposed to generate the correct weak-scale Higgs mass as well as to protect
it from the ultraviolet(UV) divergence.
In any such spontaneously symmetry breaking scenario, there are massless spin-0 (Goldstone) excitations
along flat directions of the potential that realize the underlying symmetry of the Lagrangian.
If the symmetry is gauged, these degrees of freedom are absorbed in the longitudinal modes of the newly massive vectors,
and the Goldstone equivalence theorem mandates that amplitudes for longitudinal vector bosons will be equivalent to
those of the associated Goldstone at large collision energies.
In particular, for both the Standard Model (SM) and BSM cases,
given that the $Z$ mass arises from the Higgs vacuum expectation value (VEV) $v$,
the $Z$ longitudinal modes and the Higgs boson share common couplings,
and a (near) unit ratio of $Z/h$ production is generically expected.
While there are no heavy electroweak states in the SM to decay directly to $Z$ and $h$ bosons, new states with a non-zero
electroweak charge exist in many BSM theories, and their decay branching fraction into $Z$ and $h$ may be applied 
as a very useful probe of the Higgs sector in such models.

Supersymmetry (SUSY) has been widely accepted as a viable mechanism for alleviating large UV
fermion-loop corrections to the Higgs mass.  In $R$-parity enforcing SUSY models, the lightest (LSP) and the next-to lightest
supersymmetric particle (NLSP) may be neutralinos. The NLSP may decay into the LSP along with a $Z$ or $h$.  This channel is particularly
favorable when there are no other particles in the spectrum, e.g., sfermions, appearing
between the lightest two neutralinos that may reduce the branching fraction into $Z$ and $h$. In the
minimal supersymmetric standard model (MSSM), scenarios with a bino-like LSP and higgsino-like NLSPs are quite common,
since lighter higgsinos are preferred in order to realize a smaller value of the SUSY-preserving higgsino mixing term $\mu$.
Moreover, since anomaly cancellation requires distinct $SU(2)_{\rm L}$ Higgs doublets 
($H_u, H_d$) to provide up- and down-like masses, there are two higgsino NLSPs in this case.
The $Z,\,h$ decay branching fractions of each depend sensitively on the individual neutralino mixing of $H_u, H_d$,
although the ratio of the decay branching ratios into $Z$ and $h$ is of order unity when both contributions are added,
as predicted by the Goldstone equivalence theorem. Nevertheless, the specific ratio may feature some weak
residual dependence upon the specific model parameters, particularly the ratio $\tan\beta$ of VEVs acquired by the
up and down type MSSM Higgs fields.

It is interesting to consider alternative scenarios that can impact the $Z/h$ branching ratio.  For example,
testable deviation from the MSSM could be predicted if the Higgs mixes with other fundamental scalars that couple
outside of the $SU(2)_{\rm L}$ gauge structure, such as the the singlet field $S$ in the Next-to Minimal
Supersymmetric Model (NMSSM)~\cite{Ellis:1988er, Drees:1988fc, Durand:1988rg}.  This extension 
is independently well-motivated as a solution to the naturalness problem, providing an explanation
for why the $\mu$ term might be light, of electroweak order, counter-balancing
similarly sized contributions to the $Z$ mass that emerge explicitly from the soft SUSY-breaking sector.
Specifically, the $\mu$ term arises dynamically in this context, as the VEV of a new singlet chiral supermultiplet
containing a pair of charge-parity (CP) even/odd scalars, as well as a fifth ``singlino'' neutralino.
One or both of the (pseudo)scalars may take masses above or around the 125 GeV scale, and potentially confuse
the interpretation of fermion pair mass measurements at colliders.  Since $S$ is a singlet, and does not
participate in the Higgs mechanism, its mixing into the observed Higgs scalar can reduce the Higgs
coupling to the doublet NLSPs.  Similarly, if a singlet pseudoscalar around 125 GeV emerges in
decays, it can significantly suppress the observed branching fraction into $Z$ by
enhancing the observed Higgs-like fraction.

An alternative, explicitly non-supersymmetric, spectral modification that we will entertain for the
sake of comparison and contrast involves extension of the SM using singlet-doublet fermionic (SDF)
dark matter~\cite{Cheung:2013dua}.  This type of model introduces a singlet fermion
$S$ that couples to the SM Higgs field via two heavier doublets $D_1,D_2$, allowing for
cascade decays into the same final states as the previously described NMSSM scenario.
However this scenario can potentially be distinguished from the NMSSM
by measuring the $Z/h$ ratio, since the new fermions do not alter the Higgs sector
or modify the Higgs mass.

At the LHC, the counts of $b\bar{b}$ and opposite-sign (OS) like-flavor (LF) light lepton pairs that reconstruct $h$ or $Z$ masses in a
$2Z/h+\met$+jet(s) final state can potentially be utilized in order measure the summed doublet higgsino (or
analogous heavy electroweak state) decay branching fractions when such states are discovered.
Observation of the $Z/h$ branching fraction ratio, and quantification of its compatibility with unity,
would probe the extent to which the Goldstone equivalence theorem can offer interesting constraints
on models of new physics.  The searches of interest are inherently difficult, since direct production of the
higgsino-like second lightest neutralino in the NMSSM and of the heavier neutral fermion in the
singlet-doublet extension are not generically expected to exhibit very large cross sections.
The additional jet(s) are useful for building more missing energy into
the targeted event topology, since tagging of the leptons and $b\bar{b}$ require visible decays
of $Z,h$.  This may exhaust the mass difference between the LSP and NLSP,
especially when the LSP is not massless, limiting the available missing energy. 

In Section ~\ref{sect:scenarios}, we present sample MSSM and NMSSM higgsino scenarios that are of
observational interest at the LHC, as well as a third example of a simplified fermion singlet-doublet
dark matter model~\cite{Cheung:2013dua} that leaves the SM Higgs sector (and associated implications
for the Goldstone equivalence manifest in the doublets' decay branchings) intact.
Section ~\ref{sect:sig&bkg} elaborates on experimental issues relevant to discrimination of the $Z/h$ decay ratio.
In order to ascertain the potential sensitivity of such an analysis of new heavy weak doublets at the LHC,
the collider events in both signal and background channels are generated by Monte Carlo,
and Section~\ref{sct:generation} describes the simulation setup and assumptions.
Sections~\ref{sect:4lep},~\ref{sect:2lep} detail the classification and selection
optimization applicable to the various final states.
We conclude in Section~\ref{sect:conclusions}.

\section{Benchmark Scenarios}
\label{sect:scenarios}

The first benchmark scenario to be described is an MSSM construction.
To be general, we choose a 70 GeV LSP mass, above $M_Z/2$, so that the tight invisible $Z$ decay constraints may be evaded without requiring a pure bino state. A 70 GeV LSP also evades existing LHC constraints on Higgsino pair production searches~\cite{Khachatryan:2014mma, Khachatryan:2014qwa}.
We take the NLSPs to be higgsinos that are relatively light, but that can still decay into the LSP and  the $Z,h$ bosons.
Due to its higgsino mixing, the LSP mass has to satisfy constraints arising from direct detection experiments,
e.g. LUX~\cite{Akerib:2013tjd}. We thus choose $\mu<0$ to help suppress the LSP coupling to the Higgs, as shown in
Table~\ref{tab:benchmark1}.

\begin{table}[h]
\begin{tabular}{c|ccc|ccc|c}
\hline
MSSM &$M_1$ & $\mu$ & $\tan\beta$ & $M_{\tilde{\chi}^0_1}$& $M_{\tilde{\chi}^0_2}$&
$M_{\tilde{\chi}^0_3}$& ${\xi}^{Zh}$\\
\hline
Point I &71 &-190&10 &70&198&202&3.6\\
\hline
\end{tabular}
\caption{A sample MSSM scenario with light higgsinos. The mass spectrum and decay branchings are evaluated with
{Suspect2}~\cite{Djouadi:2002ze} and the MSSM calculator as part of the MadGraph~\cite{Alwall:2011uj} package.}
\label{tab:benchmark1}
\end{table}

For point I, we assume all sfermions are heavy and decouple at leading order.  The wino is also assumed
heavy. When $\tilde{\chi}^0_2,\tilde{\chi}^0_3$ are produced at the LHC,
it is useful to consider a ratio of the decay branching into $Z$ over that into $h$, as
defined in Ref.~\cite{Dutta:2014hma},
\be 
{\xi}^{Zh} \equiv
\frac{{f_{\tilde\chi_2^0}BR(\tilde\chi_2^0\rightarrow\tilde\chi_1^0 Z)+f_{\tilde\chi_3^0}BR(\tilde\chi_3^0\rightarrow\tilde\chi_1^0 Z)}}
	{{f_{\tilde\chi_2^0}BR(\tilde\chi_2^0\rightarrow\tilde\chi_1^0 h^*)+f_{\tilde\chi_3^0}BR(\tilde\chi_3^0\rightarrow\tilde\chi_1^0 h^*)}},
\hspace{0.21cm}
\label{eq:tilded_ratio}
\ee where $f$ is the number fraction of a specific neutralino in the signal events, and
$h^*$ denotes any (pseudo)scalar at the Higgs mass. For instance, in the NMSSM $h^*$ can be either the Higgs
scalar or the singlet pseudoscalar $a_1$.
At all our benchmark points, the s-channel $Z^*\rightarrow \tilde{\chi}^0_2\tilde{\chi}^0_3$ process dominates pair
production rates and $f_{\tilde\chi_2^0}\approx f_{\tilde\chi_3^0}$, due to a suppression in the $Z\tilde{\chi}^0_i\tilde{\chi}^0_j$
coupling for $i=j$ when $\tilde{\chi}^0_2,\tilde{\chi}^0_3$ are dominantly higgsinos. 

From the Goldstone theorem, $\tilde{\chi}_i^0\rightarrow\tilde{\chi}_1^0 h\approx  \tilde{\chi}_i^0\rightarrow\tilde{\chi}_1^0 Z$
in the longitudinal $Z$ polarization. As $Z$ also has transverse polarization that couple to $\tilde{\chi}^0$, summing up the
decay branchings of $\tilde{\chi}_2^0,\tilde{\chi}_3^0$ would result in comparable yet higher decay branching into $Z$, i.e.
$\xi^{Z/h}>1$ for $f_{\tilde\chi_2^0}= f_{\tilde\chi_3^0}$. It is worth noticing that
a large $\xi^{Z/h}$ ratio can arise from a kinematic suppression when the mass gap separating the
neutralinos is relatively small, as the decay into $Z$ has a larger phase space.
If Point I is modified to feature a very light bino, $\xi^{Z/h}$ is modified to around $2$.

\medskip
The second scenario that we discuss is a singlino LSP, higgsino NLSP case in the NMSSM~\cite{Ellwanger:2009dp},
whose superpotential has the following structure,
\be
W_{\rm Higgs} \supset \lambda \hat{S} \hat{H_u}\cdot \hat{H_d} + \frac{\kappa}{3}\hat{S}^3.
\label{eq:higgspot}
\ee
An effective $\mu = \lambda\langle \hat{S} \rangle$ term is generated when the singlet field takes a VEV,
and naturalness suggest that the combination is of the order of $M_Z$.  Note that the trilinear singlet
term simultaneously generates a mass proportional to $\kappa \langle \hat{S} \rangle \equiv \kappa \mu / \lambda$.
The somewhat heavy observed Higgs mass at 125 GeV receives tree level contribution from the singlet field,
which argues for a larger singlet coupling $\lambda$ in order to help reduce dependence on multi-TeV stops and the associated fine tuning.
Interestingly, the NMSSM allows one of the scalars (and one pseudoscalar) to be very light, if it is mainly a singlet.
This extra scalar, which decays into $(b\bar{b},\tau\bar{\tau})$ at an invariant mass outside the 125 GeV window,
will be a strong indication of this model. However, if we stay in the picture that the Higgs {\it is} the lightest
of the NMSSM scalars and contains no more than 50\% singlet, the singlet cubic coupling $\kappa$ would be non-vanishing.
The correlations between a large $\lambda$, a small $\mu$, a non-zero $\kappa$, and a mass gap from the LSP greater than the Higgs
mass force the higgsino dominated $\tilde{\chi}_2^0,\tilde{\chi}_3^0$ to be at least $\sim 270$ GeV. 
We take the first (A) and third (C) benchmarks from Ref.~\cite{Dutta:2014hma}, summarized
as Points II and II$'$ in the present Table~\ref{tab:benchmark2}. 

\begin{table}[h]
\begin{tabular}{c|cccc|c|ccc|c}
\hline
NMSSM &\ $\lambda$\ &\ $\kappa$\ &\ $\mu$\ &\ $\tan\beta$\ & $m_{a_1}$ & $m_{{\tilde\chi}^0_1}$ &
$m_{{\tilde\chi}^0_2}$ & $m_{{\tilde\chi}^0_3}$& ${\xi}^{Zh}$\\
\hline
Point II&0.8 & 0.25 &220 & 2.9 & 161 & 143 & 270 & 270&2.1\\
\hline
Point II$'$ &0.8 & 0.25 &230 & 2.9 & 119 & 150 & 279 & 279&0.7\\
\hline
\end{tabular}
\caption{A pair light higgsino NMSSM benchmark points, exhibiting
over (II) and under (II$'$) production of the $Z$ relative to the $h$. }
\label{tab:benchmark2}
\end{table}

We are interested in focusing on the parameter space region where the singlet (pseudo)scalars are somewhat heavier than,
or comparable in mass to, the Higgs, and not kinematically distinguishable. For benchmark point II, the
singlet-dominated pseudoscalar $a_1$ is a fair bit heavier than the Higgs, at 161 GeV.  This kinematically prohibits 
decays of the NLSP into $a_1$, and the ratio $\xi^{Z/h}>1$ thus falls within the same range as it does in the MSSM.
For benchmark point II$'$, $a_1$ is slightly lighter at 119 GeV, which is very close to the $h$ mass and can fake the Higgs boson.
The total decay fraction into $h$ and $a_1$ will exceed
the $Z$ fraction in this case, leading to a ratio ${\xi}^{Zh}<1$ that is distinguishable from the MSSM.  

\medskip
The third scenario we consider is a non-SUSY BSM example with an unmodified Higgs sector,
specifically the singlet-doublet fermionic (SDF) dark matter model~\cite{Cheung:2013dua}.
This model extends the SM with a singlet fermion $S$ that couples to the SM Higgs field via
two heavier weak doublets $D_1,D_2$, which have $U(1)_Y$ charges of $-\frac{1}{2}$ and $+\frac{1}{2}$, respectively,
\bea
-{\cal L}_{SDF} &=  & y_{D_1} SHD_1 + y_{D_2} SH^\dagger D_2 \\
 & + & \frac{1}{2}M_S S^2+ M_D D_1 D_2. \nonumber
 \label{eq:SDF_lagrangian}
\eea 
The new fermions $S, D_1,D_2$ mix via a symmetric mass matrix,
\be 
{\cal M}=\left(
\begin{array}{ccc}
M_S & \frac{1}{\sqrt{2}}y_{D_1}v & \frac{1}{\sqrt{2}}y_{D_2}v \\
\frac{1}{\sqrt{2}}y_{D_1}v & 0 & M_D \\
\frac{1}{\sqrt{2}}y_{D_2}v & M_D & 0 
\end{array}
\right)\,.
\label{eq:mmat}
\ee

In general, the spectrum of neutral mass eigenstates $\{\chi_i^0\}$
consists of one lighter singlet-dominated state and two heavier
doublet-dominated states that behave analogously to the pair of
higgsinos in supersymmetric models, though larger couplings allow for
more mixing than is typically possible in the neutralino sector of
the MSSM.  We will focus on the parameter state where the DM
$\chi_1^0$ is light and singlet-dominated, while the two doublet-like
states $\chi_{2,3}^0$ are heavier, in order to allow for the desired decays.
For visual distinction, these fermions do not have a tilde ($\sim$)
positioned above their symbols. The mixing angle $\tan\theta\equiv
y_{D_1}/y_{D2}$ indicates the relative size of the $D_{1},D_{2}$
couplings, and $y\equiv \sqrt{y^2_{D1}+y^2_{D2}}$.

\PlotPairStack{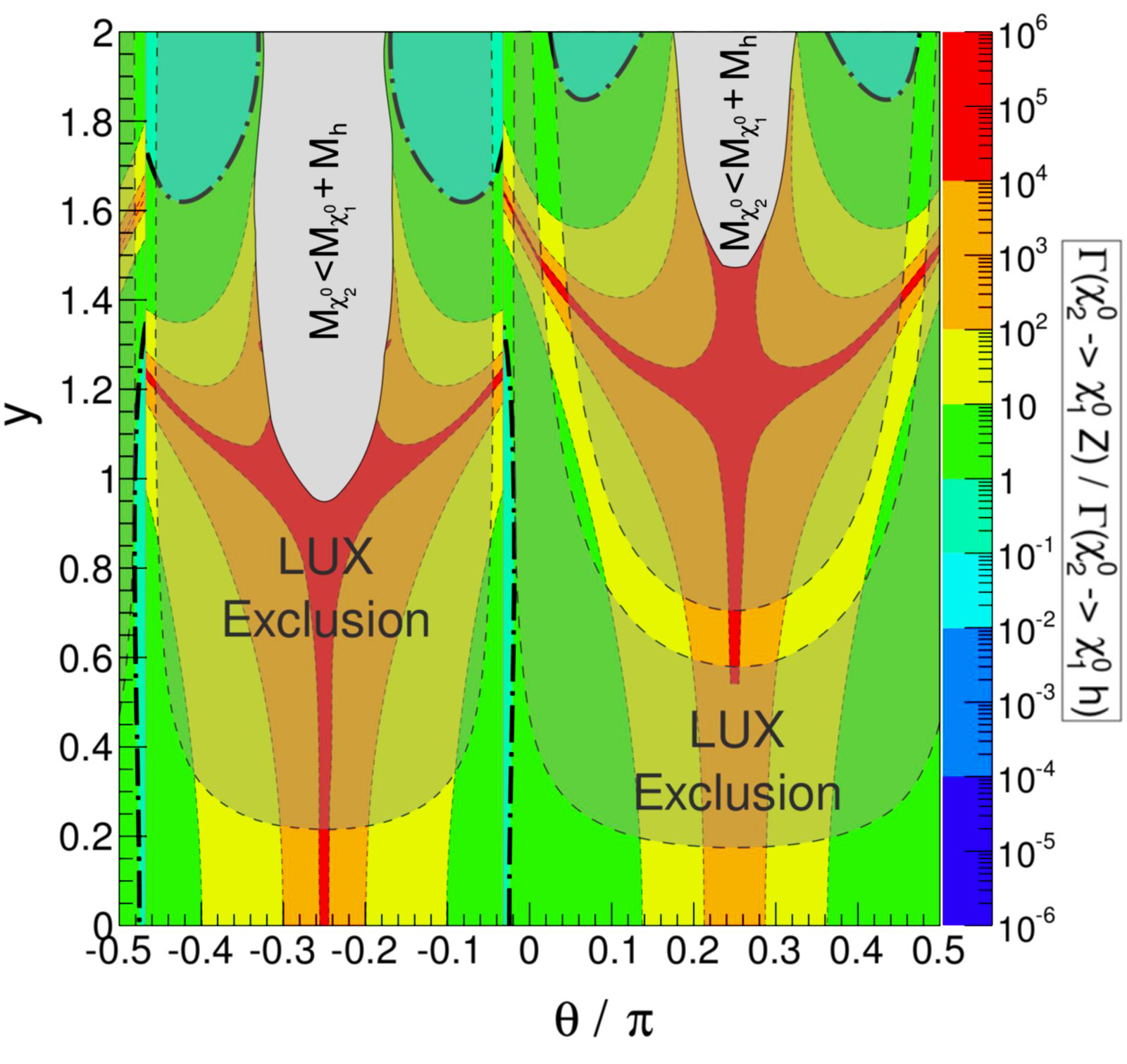}{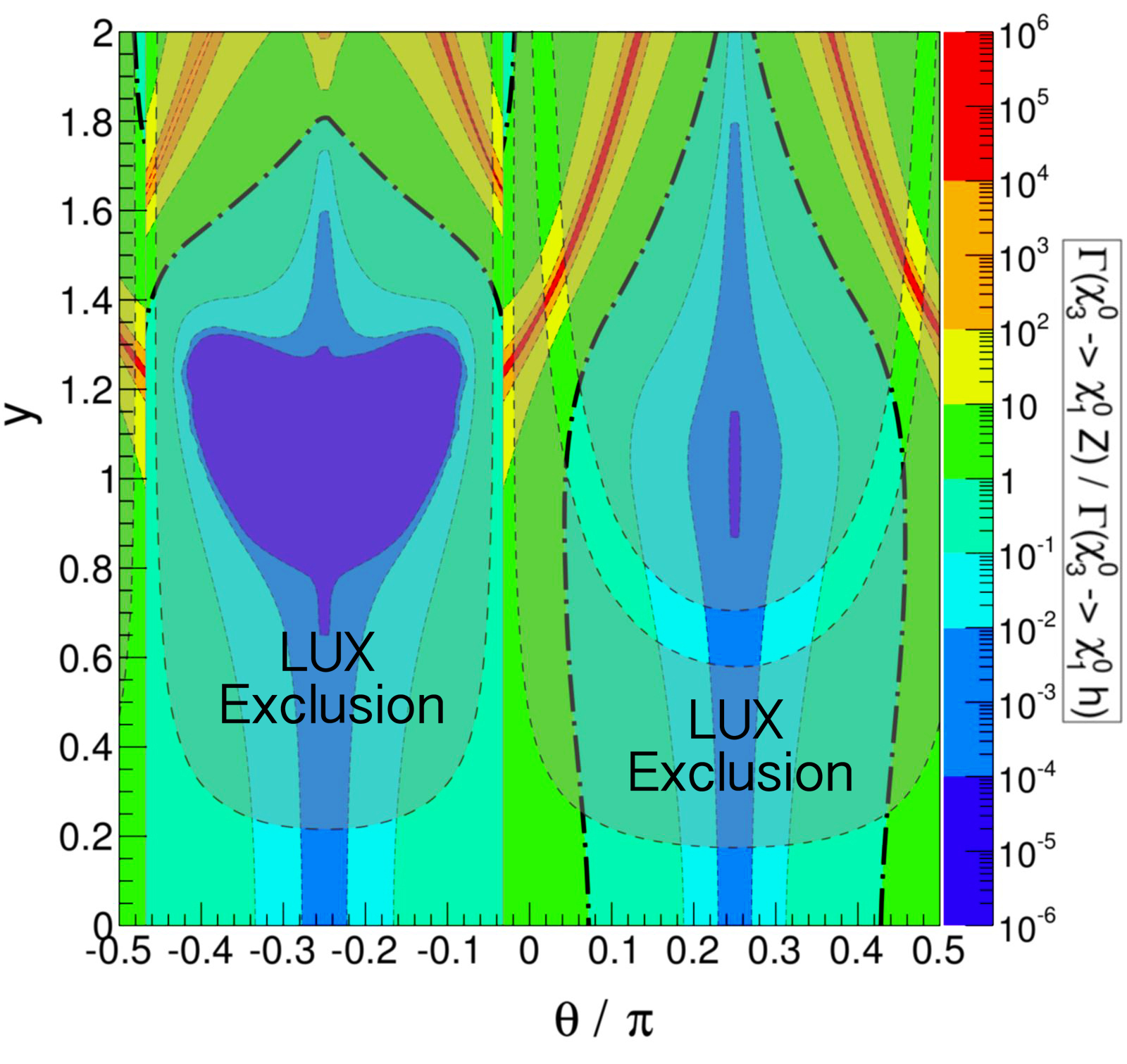}{fig:SDFratio}
              { Near symmetric branching ratio of $\chi^0_{2,3}$
                decays in the SDF dark matter model.  $M_S =
                50~\mathrm{GeV}$ and $M_D=200~\mathrm{GeV}$
                throughout.}

\begin{table}[h]
\begin{tabular}{c|cccc|ccc|c}
\hline
SDF &y &$\theta$& $M_S$& $M_D$& $M_{\chi^0_1}$& $M_{\chi^0_2}$& $M_{\chi^0_3}$ &
${\xi}^{Zh}$\\
\hline
Point III &0.4&-0.05$\pi$&72&189&70&201&203&3.0\\
\hline
\end{tabular}
\caption{Benchmark SDF dark matter point.}
\label{tab:benchmark3}
\end{table}

Neglecting any potential loop-order correction to the Higgs mass, the
terms in Eq.~\ref{eq:SDF_lagrangian} leave the SM Higgs sector
unchanged, and the Goldstone equivalence theorem predicts similar
branching fractions in the $\chi^0_{2,3}\rightarrow \chi^0_1,Z/h$
decays.  Fig.~\ref{fig:SDFratio} show the $Z/h$ ratios associated with
$\chi^0_{2,3}$ decays are approximately symmetric, up to corrections
from the kinematic and mixing differences. The $ySHD$ term plays a
central role in providing the NLSPs comparable decay width into $Z,h$
(see Appendix~\ref{app:Zh_in_SDF} for details).
The doublet component in the DM leads to DM-nucleon scattering via $Z$ boson, 
and the shaded regions in Fig.~\ref{fig:SDFratio} denote the
parameter space corresponding to a large DM-nucleon scattering cross section
that is ruled out by LUX~\cite{Akerib:2013tjd}.
The gap between the two shaded areas in the right half ($\theta > 0$)
of each panel identifies an allowed region of parameter space where the
lightest fermion, i.e. the DM candidate, mass becomes sufficiently light 
that LUX loses sensitivity and cannot rule out the scenario.
The discontinuity at
$\theta \approx -0.03\pi$ occurs because the pair of doublet-like
states become Dirac once more at that point, resulting in a phase
shift in the rotation matrix across the transition.
We refer to Ref.~\cite{Cheung:2013dua} for the nucleon scattering cross-section
calculation, and other phenomenological studies of this model.

Table~\ref{tab:benchmark3} provides an SDF benchmark (Point III). The
mixing angle is chosen to be $\theta = -0.05\pi$, where the LSP-nucleon
scattering is suppressed and the coupling $y$ can take relatively large
values.  Similar to the MSSM, the Point III also gives rise to a
higher decay branching into $Z$, as predicted by the Goldstone
equivalence theorem. This shows that, in the simplified picture of weak
doublet fermion decays, the unaltered Higgs sector also yields a ratio 
$\xi^{Z/h}>1$; this stands in contrast to the situation where a new field
masking the Higgs is additionally present, as in the case of NMSSM.
However, similar final states and a similar $\xi$ ratio make this model
difficult to distinguish from other constructions like the MSSM, underscoring
a need for caution in the interpretation of results that may be mutually
associable with a degeneracy of underlying structures.
In this case, more knowledge about the model's particle
spectrum will be needed.

\section{Experimental Considerations}
\label{sect:sig&bkg}

Pair production of the various heavy weak doublet candidates identified with the aforementioned
scenarios constitutes a potentially viable search channel at the LHC.
The decay of these doublets and the subsequent decay of the associated $Z/h$ boson products
lead to final states that contain leptons and $b$-tagged jets, plus some amount of missing transverse
energy $\met$, and (optionally) additional jets.

The $Z$ boson will decay dominantly to hadrons (70\%), including a 15\% share to just the $b\overline{b}$ final state.
Invisible decays account for 20\% of the branching, and the final 10\% is shared among the three lepton
pair production modes $\ell\ell$.  There is no intrinsic preference for or against $\tau$ production
in this mode, but we will focus on the selection only of light lepton ($e,\mu$) flavors because they have
a much higher detection efficiency, and a lower fake rate.  The Higgs boson $h$ will be reconstructed from
its decay into  $b\overline{b}$, at a large branching which is under correction from the size of $\tan\beta$
in supersymmetric models. If a $\sim 125$ GeV NMSSM singlet pseudoscalar emerges, it dominantly decays into $b\bar{b}$
with a branching near 100\%. 

Light leptons will not be produced directly at any appreciable rate by the decays of the Higgs, although there may be
leptonic decays arising from its direct decay products, with light opposite-sign mixed-flavor pairs (along with the associated
missing neutrinos) represented in the final state at a typical branching of about 1\% each (for a SM Higgs) via
the $WW^*$ and $\tau\tau$ channels.  These rates may be further discriminated from the direct decays of the $Z$ by demanding same
flavor combinations with kinematic reconstruction of the parent mass.

From the decay of mixed $Z/h$ pairs, the targeted final states will therefore correspond to 4
leptons $\ell\ell\ell\ell$, 4 $b$-tags $b\overline{b}b\overline{b}$, or a mixed state $\ell\ell b\overline{b}$.
The lepton production channels will be associated with $Z$ boson decays, and the $b$-tag production
channels will be dominantly associable with decays of the Higgs (and Higgs-like states).
Two out of the three described signals must be measured in order to
ascertain the parent doublets' total decay branching ratio into $Z$ and $h$.

The 4-lepton channel has a substantial SM background of vector bosons plus Jets,
where the vector boson, e.g. the $Z$, decays leptonically and
jet mismeasurement provides a source of missing energy.
Contributions include $t$-channel vector pair production,
and $s$-channel single production of a vector resonance with
one leg of the ensuing leptonic decay further radiating a second off-shell vector;
our simulation reflects a strong contribution from the former.
These backgrounds can be efficiently controlled by a $\met$ cut and by invoking
variables designed to discriminate against the spurious appearance
of $\met$ that is faked by jet mismeasurement.
Profitable selection alternatives will be discussed in Section~\ref{sect:4lep}.

By comparison, the 2 and 4 $b$-Jet channels are mainly affected by a more severe SM $t\bar{t}$ background,
where the two natural $b$ quarks from top decay may be readily accompanied by
additional additional jet-faked $b$'s.  Given the crowded final state, the rate of fake $b$-Jets
is non-negligible, and this background remains quite severe even after a $b\bar{b}$ invariant mass window cut is imposed.  
Compared to the 4$\ell$ final state, channels involving fewer leptons (and correspondingly higher $b$-tag requirements)
will thereby turn out to have substantial disadvantages with regards to detection efficiency,
as will be shown in the analysis of specific event selection alternatives in Section~\ref{sect:2lep}.
If the LSP is massless, Run I of the LHC sets limits in the 4$b$ channel,
but there are no existing limits for a massive LSP scenario.

When the mass gap between the decaying doublets and the LSP is close to (or less than) 125 GeV, the decay into $h$ may
become kinematically suppressed, leading to $Z$ dominated final states with the observable consequence that ${\xi}^{Zh} > 1$.
Additionally, even when decay into $h$ is allowed, it can still be very difficult in this regime to boost an appreciable quantity of $\met$.
It is helpful then to tag on initial state radiation (ISR) jets in order to boost the overall $\met$ of the visible system,
but 2 ISR jets are observed in simulation (for the Point II benchmark specifically) to cost a half magnitude order in production cross section.

Further details regarding the mode of simulation, the specific mechanisms available for
controlling various backgrounds, and the expected visibility of the three targeted final state
signal topologies at the $\sqrt{s} = 14$~TeV LHC are provided in the following sections.
We will not attempt a detailed extraction of the ${\xi}^{Zh}$ ratio from simulated collider data
in this work, but will instead direct attention toward the preliminary task of establishing the
signal. We will reference NMSSM benchmark II for concreteness, and comment on the extrapolation
of results to other benchmark scenarios.

\section{Event Generation and Selection}
\label{sct:generation}

Signal and the standard model (SM) background Monte Carlo event samples,
including parton showering and fast detector simulation, are generated via the standard
{\sc MadGraph5}/{\sc MadEvent}~\cite{Alwall:2011uj}, {\sc Pythia}~\cite{Sjostrand:2006za}, {\sc Delphes 3}~\cite{deFavereau:2013fsa} chain.
{\sc MadEvent} is configured, in conjunction with {\sc Pythia}, to use MLM matching.
The {\sc Delphes 3} detector simulation employs a standard LHC-appropriate parameter card, with jet
clustering performed using the anti-kt algorithm.
Selection cuts and computation of collider observables are implemented within {\sc AEACuS~3.15}~\cite{Walker:2012vf,aeacus}
using the instructions in Card~\ref{CARD:GENERAL}, as exhibited in Appendix~\ref{app:card}.
At the preselection stage jets (including $b$-tagged jets) are accepted with a transverse momentum $\pT > 30$~GeV,
up to a pseudorapidity magnitude of $|\eta| < 2.5$ (although wide jets $|\eta| < 5.0$ are employed
for limited purposes such as counting of single-track jets).
Leptons, including hadronic taus, are accepted with $\pT > 10$~GeV and $|\eta| < 2.5$.
Light leptons ($e,\mu$) are required to maintain a mutual isolation of $\Delta R > 0.3$.

\bgroup
\def\arraystretch{1.3}
\begin{table}[!htp] 
\caption{Matched production and residual effective cross section (fb) at the LHC14
are tabulated for the three targeted final state event topologies, reported individually for
the $\ttbar+$Jets and $VV+$Jets backgrounds, as well as the benchmark Point II NMSSM signal.}
\label{tab:cats}
\begin{center}
\begin{footnotesize}
\begin{tabular}{c c c c c c}
\hline 
Selection & $\ttbar+$Jets & $VV+$Jets & $W/Z+$Jets & Signal \\
\hline
Matched Production \quad & 613,000 & 150,000 & $2.27\times10^8$ & 53 \\
\hline
Cat I ($4^+e/\mu$, $0^+$B's) & 0 & 11.6 & 0 & 0.037 \\
\hline
Cat II ($2\text{--}3~e/\mu$, $2^+$B's) & 3590 & 12.8 & 62.6 & 0.130 \\
\hline
Cat III ($0\text{--}1~e/\mu$, $4^+$B's) & 1430 & 6.43 & 147 & 0.114 \\
\hline
\end{tabular}
\end{footnotesize}
\end{center}
\end{table}
\egroup

\bgroup
\setlength{\tabcolsep}{8pt}
\def\arraystretch{1.3}
\begin{table*}[!htp] 
\caption{Summary of optimized secondary event selections employed
for each of the three targeted final state event topologies.
Also presented are the (sequential flow) percentages cut of residual events
for the background (B) and signal (S) respectively,
where B invokes the unified SM components $\ttbar$+Jets, $VV$+Jets, and $W/Z$+Jets.
Statistics for the baseline topology of each event selection category were presented in Table~\ref{tab:cats}.}
\label{tab:flow}
\begin{center}
\begin{footnotesize}
\begin{tabular}{c c | c c | c c }
\hline 
Cat I ($4^+e/\mu$, $0^+$B's) & \% (B,S) & Cat II ($2\text{--}3~e/\mu$, $2^+$B's) & \% (B,S) & Cat III ($0\text{--}1~e/\mu$, $4^+$B's) & \% (B,S) \\\hline 
$\tau$ Veto &(0.4,0.7)& $\tau$ Veto &(1.1,1.2)&  2 Leading B-Jets  &(50,28)\\\hline 
$b$-Jet Veto &(0.4,1.5)& $1^+$ Hadronic $Z/H$ &(61,21)& $e/\mu$ Veto &(12,0.8)\\\hline 
$\met/\sqrt{H_{\rm T}} > 6.0~{\rm GeV}^{1/2}$ &(100,64)& $1^+$ Leptonic $Z$ &(95,27)& $\tau$ Veto &(5.7,2.4)\\\hline 
&& 1-Track Jet Veto &(1.3,1.6)& $2^+$ Hadronic $Z/H$ &(58,34)\\\hline 
&& $\Delta R < 2.0$ &(56,40)& $6^+$ Jets Veto &(52,13)\\\hline 
&& $\met/\sqrt{H_{\rm T}} > 3.0~{\rm GeV}^{1/2}$ &(57,28)& 1-Track Jet Veto &(2.4,1.4)\\\hline 
&&&& $\met/\sqrt{H_{\rm T}} > 3.0~{\rm GeV}^{1/2}$ &(72,35)\\
\hline
\end{tabular}
\end{footnotesize}
\end{center}
\end{table*}
\egroup

\PlotSingle{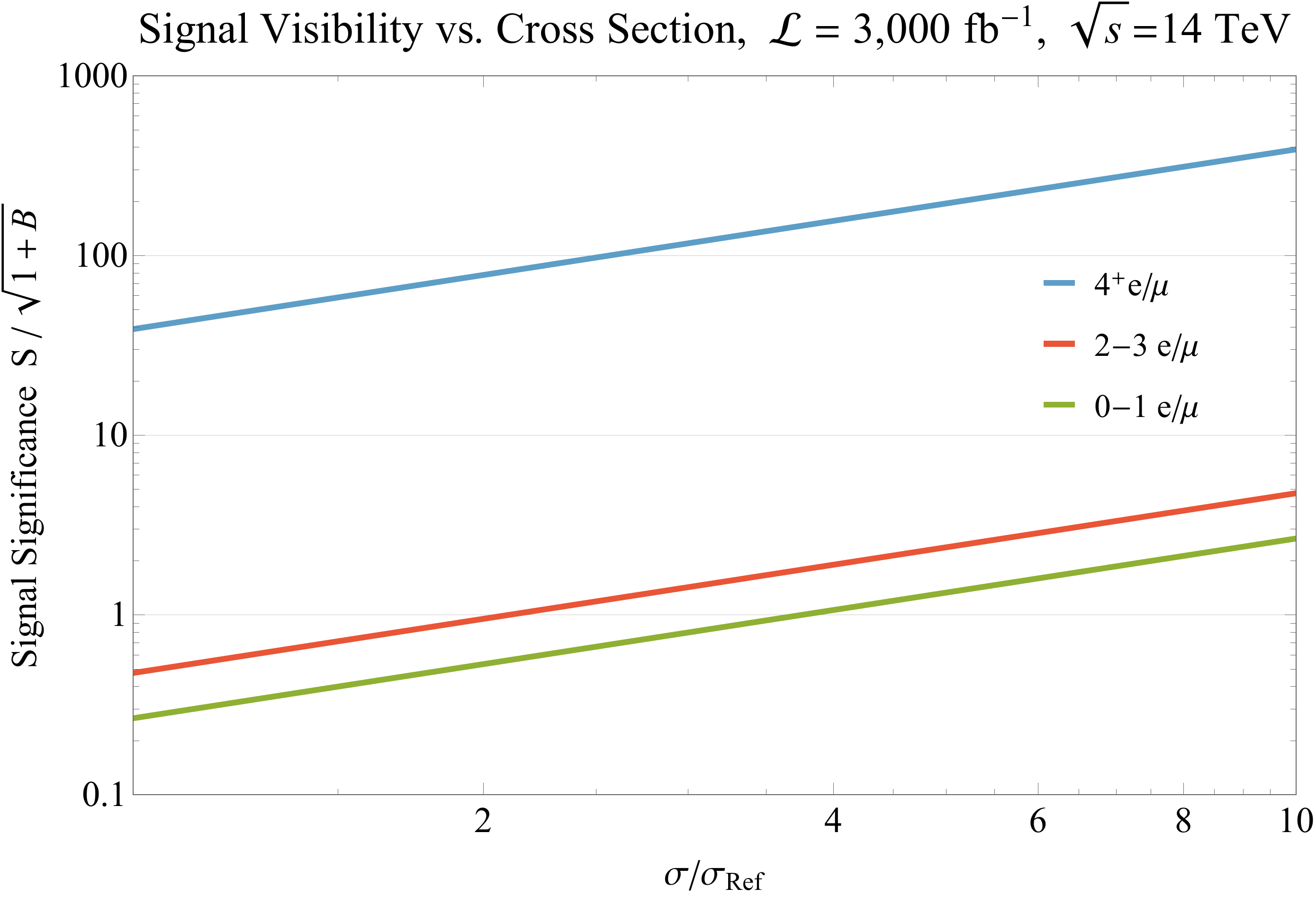}{3.4in}{fig:signif}
{
The signal significance metric $S/\sqrt{1+B}$ is projected
for each of the targeted final state topologies at a
luminosity of 3,000 ${\rm fb}^{-1}$ at the $\sqrt{s} = 14$~TeV LHC,
using statistical errors on the simulated background estimation only.
Collider modeling is based on NMSSM benchmark II, which features 270 GeV
higgsinos and a $Z/h$ ratio of $\xi^{Z/h} = 2.1$.  The horizontal 
axis represents numerical scaling of the production cross section
relative to this benchmark.  The optimized event selections employed
are summarized in Table~\ref{tab:flow}.  Supplementary cuts on $\met$
are not considered here, although this could become a favorable strategy
at very large luminosity or cross section for the
the middle $2\ell 2b$ scenario, as elaborated in Section~\ref{sect:2lep}.
Notice that the scaling of the post-cut $\sigma$ depends on both the higgsino production
cross-section and the neutralino decay branching into the specific final state categories.
Particularly at large luminosity, systematic errors in the background will likewise be important,
summed in quadrature with the statistical fluctuation of the background.
}

Background candidates simulated here are the inclusive production of $\ttbar$ with 0-2 jets, the
inclusive production of two vector bosons (meaning $W$ and/or $Z$) with 0,1, or 2 Jets,
as well as single $W$- or $Z$-boson production including 0-4 initial state Jets.
The single vector channels exhibit very large production cross sections,
around two orders larger than the corresponding $\ttbar$ background, and the approximately 15 
million events considered for each case remain a substantial under-sampling.
Approximately 3 million events were sampled for the $VV$+Jets background, corresponding to
around 20 ${\rm fb}^{-1}$ of integrated luminosity.
More than 60 million events were sampled for the $\ttbar$+Jets background, corresponding to
around 100 ${\rm fb}^{-1}$ of integrated luminosity.
For signals, we have simulated around 25 million events for the NMSSM benchmark II point,
likewise inclusively considering 0-2 Jets, which corresponds to
an integrated luminosity approaching half a million events per femtobarn.

Signal events are pre-classified into three non-overlapping categories based on the count of light
leptons ($e,\mu$), along with a complementary threshold for the count of of heavy-flavor ($b$-tagged) jets,
as stipulated at the bottom of the Card~\ref{CARD:GENERAL} instructions.
Category I contains at least 4 leptons, but has no $b$-tagging requirement. Category II contains either 2 or 3
leptons, and at least 2 $b$-Jets. Category III contains either 0 or 1 leptons, and at least 4 $b$-Jets. There is
significant attenuation of both signal and background by these preliminary topological cuts, as
demonstrated in Table~\ref{tab:cats}.  Category I is intrinsically low-background, and the signal already competes well here, being of the same
magnitude order as the isolated $\ttbar+$Jets and $VV+$Jets components. Categories II and III are
dominated by the $\ttbar+$Jets background, which shall prove quite difficult to reduce while retaining
any appreciable portion of the already meager signal.  Additionally, we preemptively summarize in Table~\ref{tab:flow}
the supplementary event selection optimizations and cut flow for each of these event categories,
which will be established in the following sections.

It should be emphasized before proceeding that the baseline NMSSM matched production cross section ($\sim 50$ fb) provided 
in Table~\ref{tab:cats} is for the particular neutralino mass $M_{n_{2,3}} \sim 270$ GeV associated with benchmark II,
which has been selected as the default scenario for our collider study.
The cross section can be a quite a bit larger when the associated doublets are lighter, as can occur in the MSSM and in the SDF models.
If the relevant mass scale, e.g. for the higgsino-type MSSM neutralino or for the SDF doublet, is reduced from
270 GeV to around 200 GeV, an increase in the higgsino pair-production cross section
by a factor of order five can generically be expected~\cite{Han:2013usa,Delannoy:2013ata}.
In Fig.~(\ref{fig:signif}) we preemptively summarize the optimized visibility of each signal
region at a luminosity of 3,000 ${\rm fb}^{-1}$ as a rescaled function (statistical errors only) of the post-cut cross section.
Kinematic cut efficiencies are expected to be less affected by scaling when the mass gaps between the LSP and NLSP(s)
remain as represented by the benchmarks.  The scaling is affected by both the higgsino production rate and the
ratio $\xi^{Z/h}$ that determines the neutralinos' branching into each final state category.
Benchmark II, with $\xi^{Z/h}=$2.1, thus inherits a greater share of leptonic final states.
For a model with a low $\xi^{Z/h}$, like benchmark II', a higher branching ratio
into $b\bar{b}$ would then enhance the $bbll$ significance (relative to $4l$).
The 4-lepton (Category I) signal region is found to be highly visible, whereas
the mixed lepton plus $b$-Jet signal region (Category II) is
conditionally visible, and the 4 $b$-Jet signal region (Category III) projects low visibility.
The applicable event selection strategies in each event category are developed in detail in the subsequent sections.

\section{Refining the 4 Lepton Signal}
\label{sect:4lep}

A natural final state to target for models similar to the NMSSM benchmark under consideration is
the Category I 4-lepton topology.
This final state has been carefully studied at the LHC~\cite{Aad:2014wra}.
The question of whether it is possible to improve the discrimination of signal from background
is investigated in the present section. To begin, a sequence of plots is shown that compare the
normalized event shape distributions of the signal and background for several observables.
All plots have been generated with the {\sc RHADAManTHUS 1.2}~\cite{aeacus} software package.
The single vector backgrounds have been integrated with the di-boson production channel.
Moderate bin smoothing is employed.

Events featuring $\ttbar$+Jets production are generally unable to legitimately
yield more than two leptons, and simulation suggests that the likelihood of
this background faking the targeted four lepton final state may likewise be discounted. 
The leading vectors plus jets background is capable of producing this event topology
directly, although the branching fraction for $Z$ to $4\ell$ is at the $10^{-6}$ order~\cite{Aad:2014wra};
the four-lepton requirement essentially rules out final states with neutrinos, and any missing
transverse energy associated with this production mode will typically arise from measurement error.
This observation suggests that collider variables designed to root out fake missing energy
signals may be very helpful here, such as the $\met$-jet angular difference
$\Delta\phi$~\cite{ATLAS-CONF-2012-033} (applied as the minimal azimuthal separation between the
missing transverse energy and the leading and $b$-tagged jets), and the missing energy significance
$\met/\sqrt{H_{\rm T}}$.

\PlotPairWide{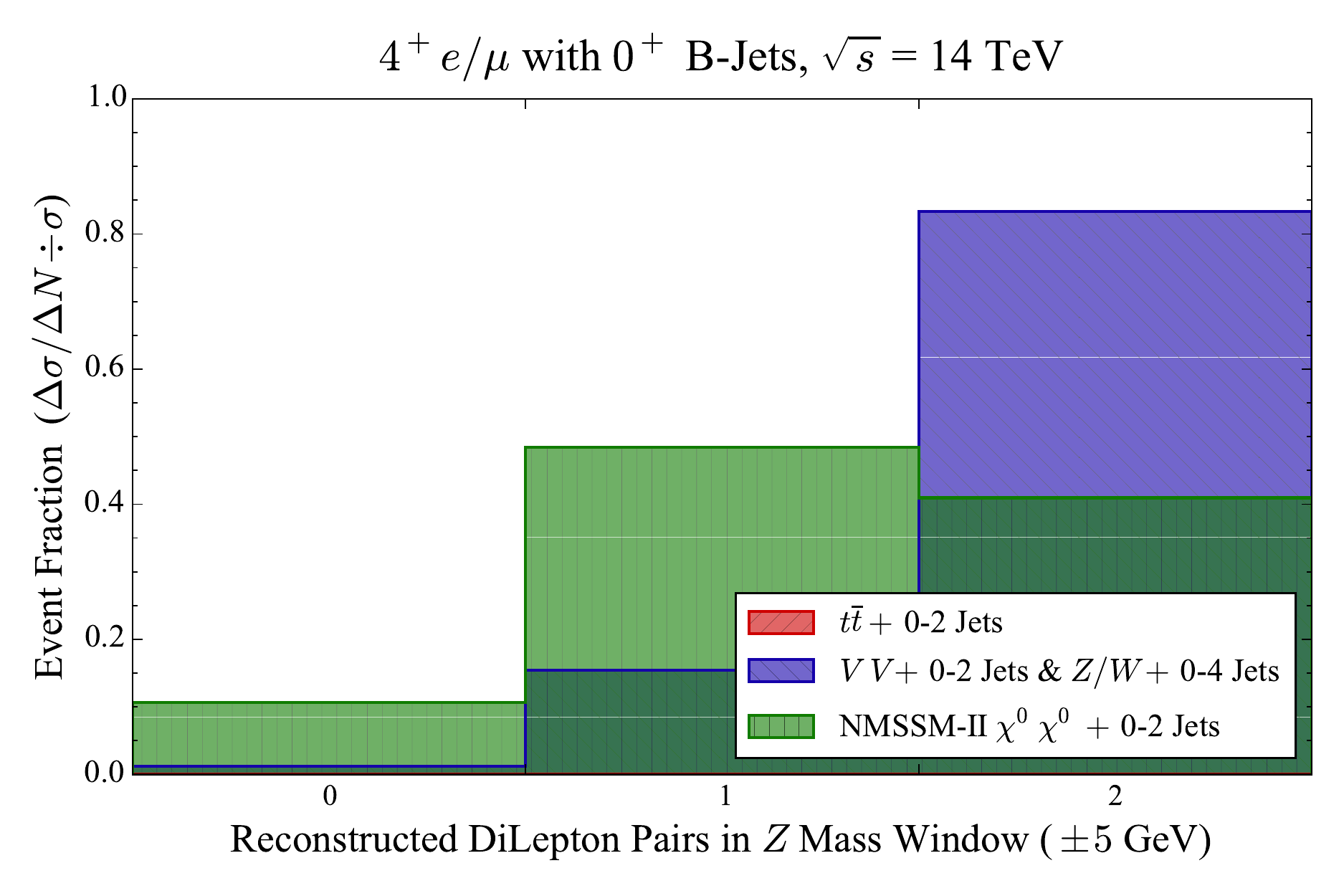}{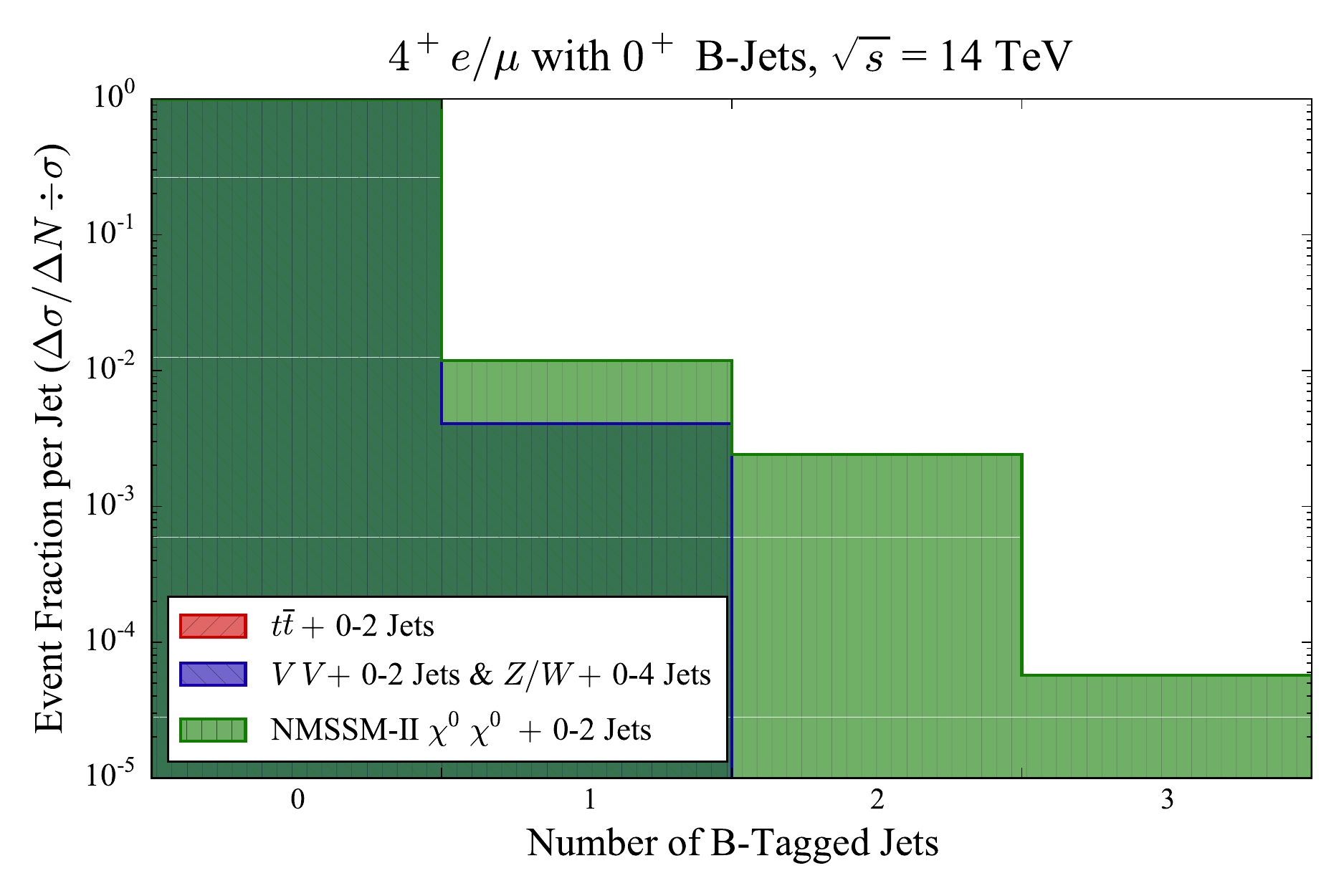}{fig:catI_LEP}
{
Signal and background event shapes are compared for the
final state topology with $4^+$ light leptons (category I).
The $\ttbar$ background component delivers no appreciable contribution to this final state.
Left: Over 90\% of the signal features a reconstructed OS-LF dilepton in the $Z$-boson mass
window ($92\pm5$)~GeV.  About half of the signal
reconstructs precisely a single $Z$, whereas almost 85\% of the
unified vector backgrounds are actually observed to reconstruct two pair.
Right: Neither the signal nor the dominant $VV+$Jets background are likely
to be tagged for a $b$-Jet at beyond the percent level, although likelihood
for the signal is somewhat greater by comparison.
}

\PlotPairWide{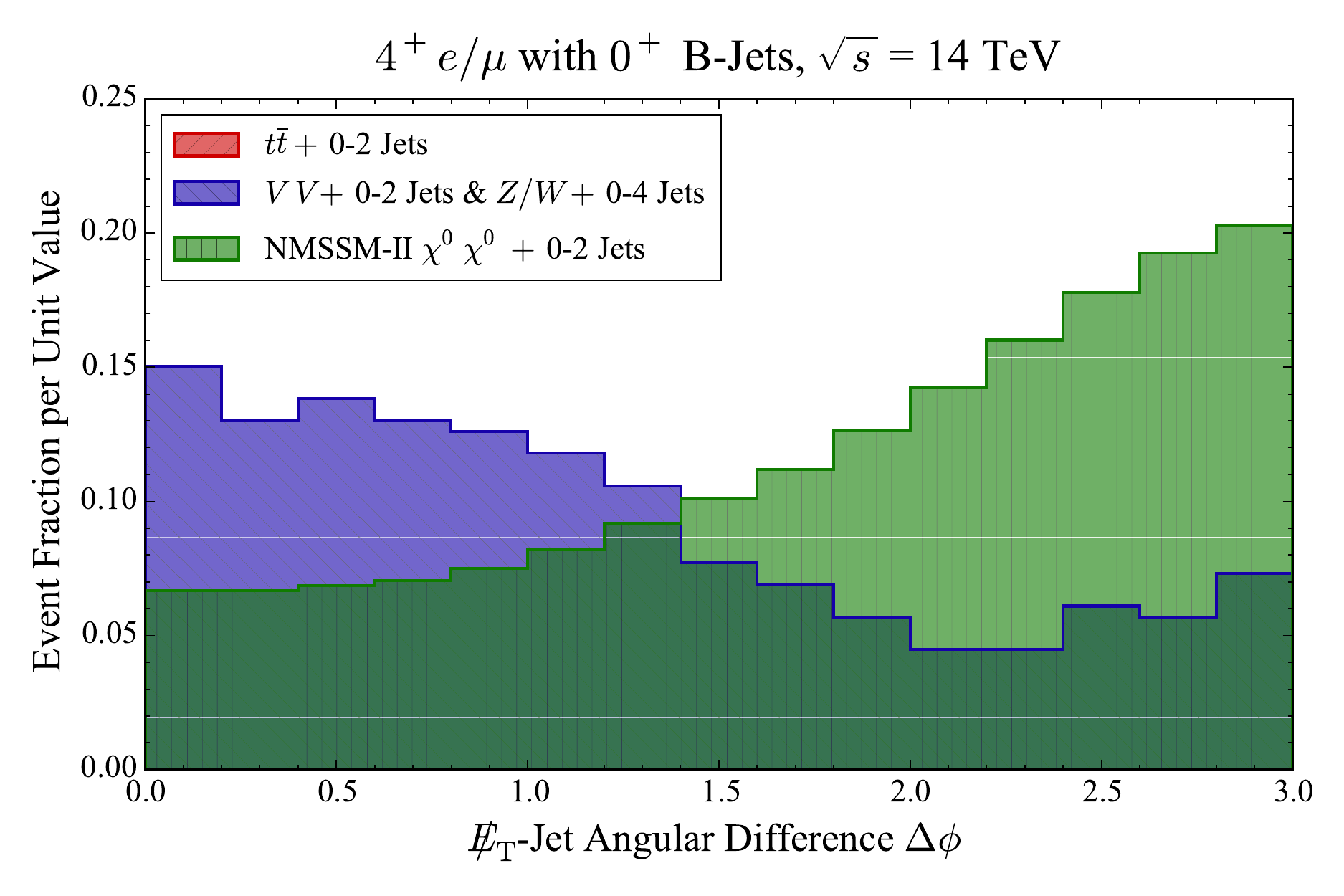}{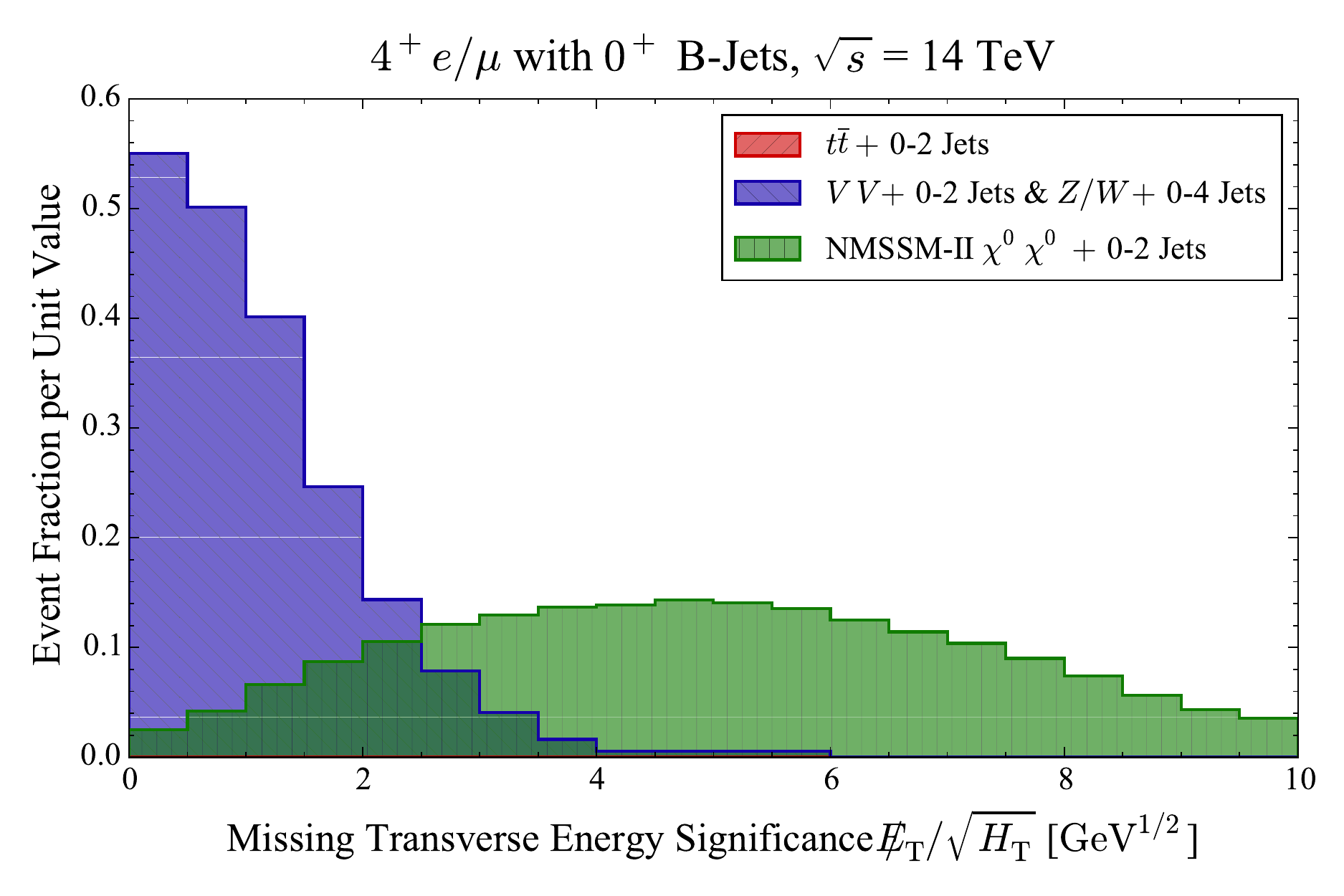}{fig:catI_RHR}
{
Signal and background event shapes are compared for the
final state topology with $4^+$ light leptons (category I).
The $\ttbar$ background component delivers no appreciable contribution to this final state.
Left: The leading vectors plus jets background relies on jet mismeasurement in order to
generate missing energy, and the $\met$ azimuthal direction is thereby here observed to be
much more strongly correlated (smaller $\Delta\phi$) with the direction of a single hard jet
than is the case for the signal's legitimate missing energy.
Right: Likewise, the quantity of missing transverse energy observed in signal is generally
a much more substantial multiplier of the estimated uncertainty $\sqrt{H_{\rm T}}$ in the
hard event scale.
}

\PlotPairWide{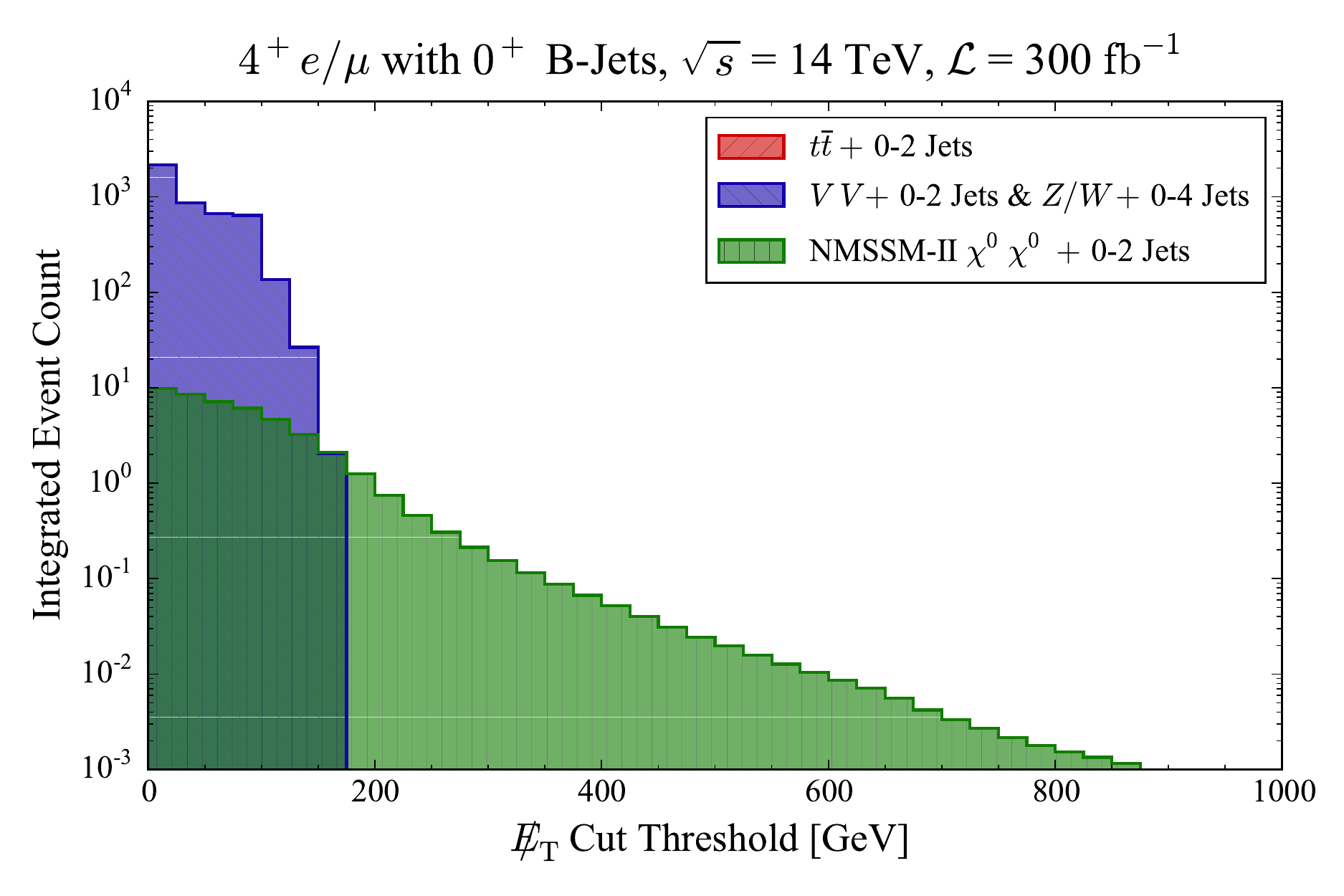}{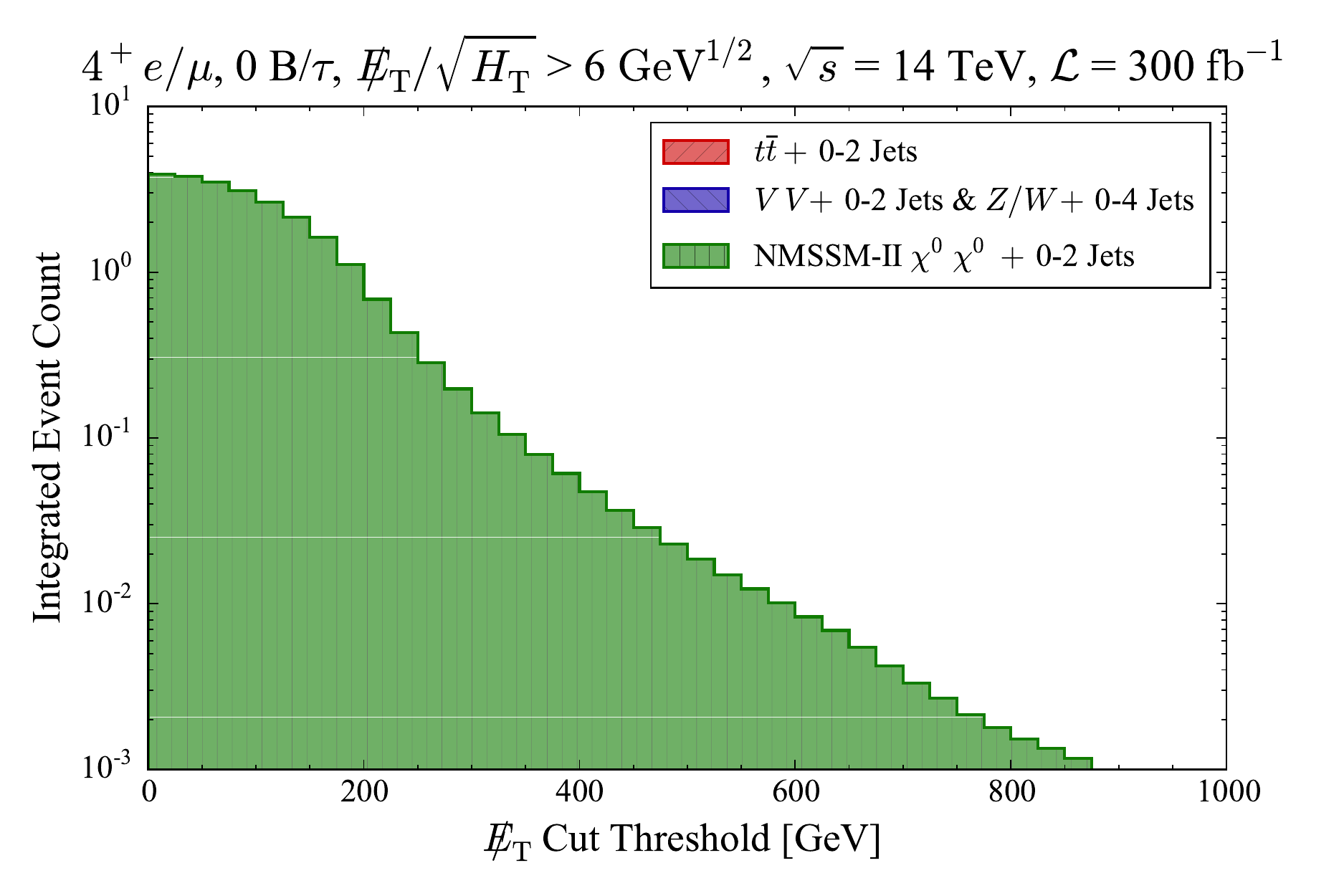}{fig:catI_MET}
{
Signal and background integrated event counts are compared for the
final state topology with $4^+$ light leptons (category I) at a luminosity of 300
events per femtobarn as a function of the missing transverse energy $\met$ cut threshold.
Left: The raw event categorization is intrinsically low background, although a weak
signal may still struggle to compete at low missing transverse energy.
Right: Enacting the secondary event selections ($\met/\sqrt{H_{\rm T}} > 6.0~{\rm GeV}^{1/2}$,
0 $\tau$, 0 B's) suggested by Figs.~(\ref{fig:catI_LEP},\ref{fig:catI_RHR}) preferentially
suppresses the background, to a point approaching elimination. There is then no residual
necessity for large missing energy, although neither is a modest cut in the vicinity
of $\met > 100$~GeV strongly disfavored.
}
 
In Figs.~(\ref{fig:catI_LEP}), the signal is differentiated from the
leading $VV$+Jets background as more likely to feature precisely one candidate $Z$-boson,
defined as an OS-LF dilepton pair with an invariant mass of $92\pm5$~GeV (left panel).
More specifically, the unified vector backgrounds are more likely, by about six times,
to contain two pair rather than one, whereas the signal is observed to contain
(0,1,2) reconstructible $Z$-bosons in approximately (10,50,40) percent of events, respectively. 
Neither the signal nor leading background are likely beyond the percent level to feature $b$-Jets,
although the signal events are slightly more so likely (right panel).
Similarly, the signal generally contains no hadronic taus.
It is also noted that the signal production cross section for 2 jets is smaller by a factor of
almost four than the matched inclusive cross section with 0-2 jets; the signal is likewise
not very jetty in character.

In Figs.~(\ref{fig:catI_RHR}), the signal and vectors plus jets background are observed
to behave consistent with the respective expectations for a legitimate and
measurement-induced missing energy source. Specifically, the background favors small
$\Delta\phi$, with $\met$ well-aligned to a hard jet, whereas the signal is characterized
by larger $\Delta\phi$ angles, indicating less correlation between the $\met$ and hard
jet directionality (left panel); this variable is best suited for application to
signals such as the one under consideration that are not overly jetty.
Likewise, the quantity of missing transverse energy observed in signal is generally
a much more substantial multiplier of the estimated uncertainty $\sqrt{H_{\rm T}}$ in the hard event scale (right panel).

Based upon these observations, supplementary event selection may be performed,
corresponding to a requirement of $\met/\sqrt{H_{\rm T}} > 6.0~{\rm GeV}^{1/2}$;
this cut essentially eliminates the vectors plus jets background in our simulation,
while retaining approximately 40\% of the signal.  Additional discrimination may
be achieved by requiring no more than one OS-LF
di-lepton invariant mass reconstruction within 5 GeV of the $Z$ boson,
or by requiring $\Delta\phi > 2$, although the original selection is
sufficient in our simulation to squelch background, while retaining the
largest fraction of an already tenuous signal strength. 
It will typically do no harm to additionally impose a veto on $b$-Jets
and hadronic taus; although this does not strengthen discrimination against the vector
backgrounds, it may further harden the exclusion against fakes from channels such as $\ttbar$+Jets.
In fact, it will suppress the signal by no more than about ten percent
to rule out events with more than one jet of any type.

In Figs.~(\ref{fig:catI_MET}), the absolute event counts attributable to
the signal and background components are compared as a function of the
missing transverse energy $\met$ cut threshold for an integrated luminosity
of 300 ${\rm fb}^{-1}$; the left-hand panel corresponds to the raw category I pre-selection,
whereas the described secondary event selections are enacted in the right-hand panel.
In this case, the residual discrimination power of the missing transverse energy
variable is apparent, with the background rate observed to drop below the signal rate
in the vicinity of 175 to 200 GeV.  However, the absolute signal rate remains rather low
at the studied luminosity in this scenario, at close to the unit level.
The effect of enforcing a hard cut on the significance estimator
$\met/\sqrt{H_{\rm T}} > 6.0~{\rm GeV}^{1/2}$ is apparent in the right-hand
panel, where backgrounds are eliminated while retaining about four signal
events at the simulated luminosity and cross section.
Even after a mild cut on the missing transverse energy, no greater than about 100~GeV,
the projected signal significance is in a favorable range close to 4.

\section{Refining the 2-3 Lepton Plus $2^+$ B-Jets and $4^+$ B-Jet Signals}
\label{sect:2lep}

\PlotPairWide{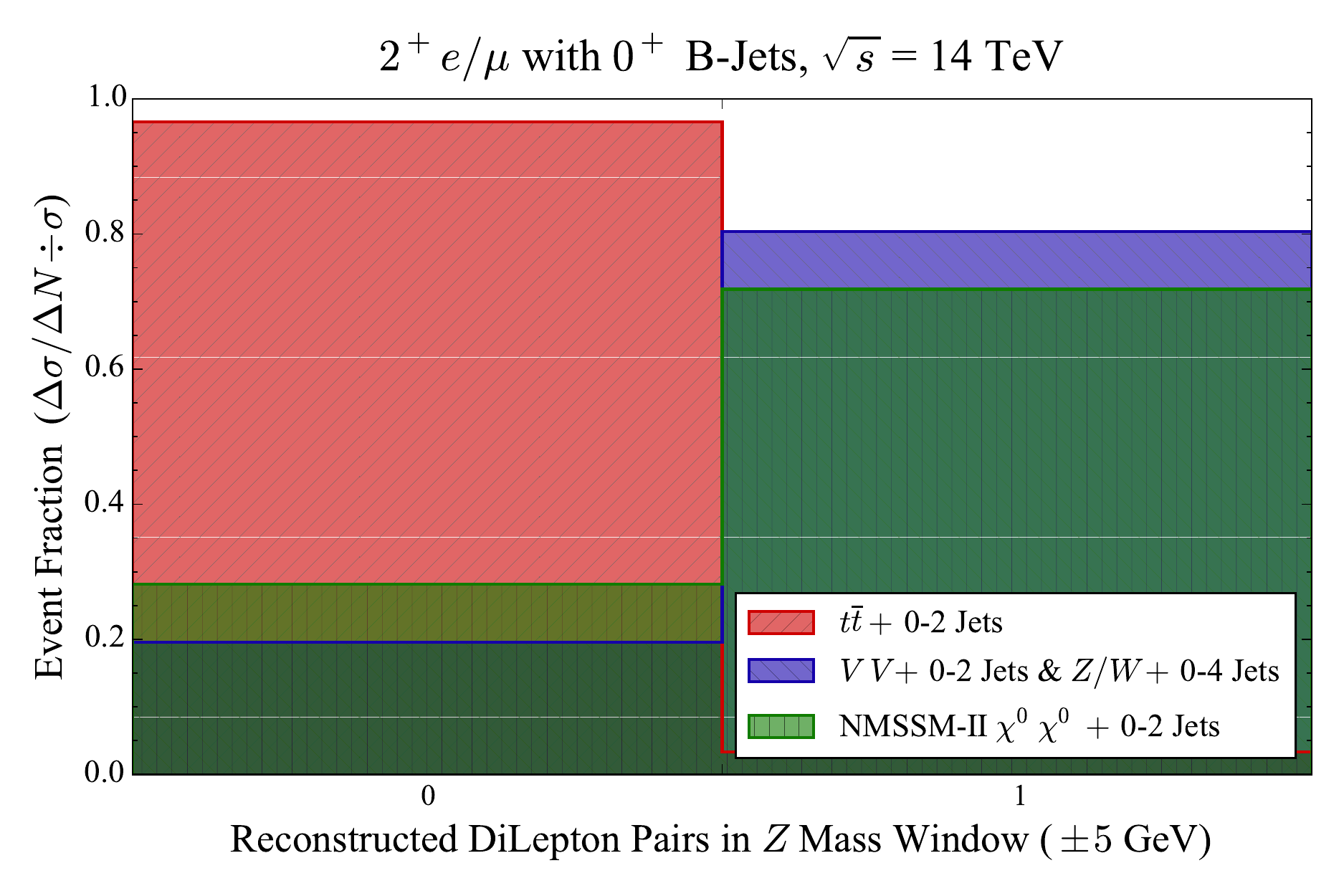}{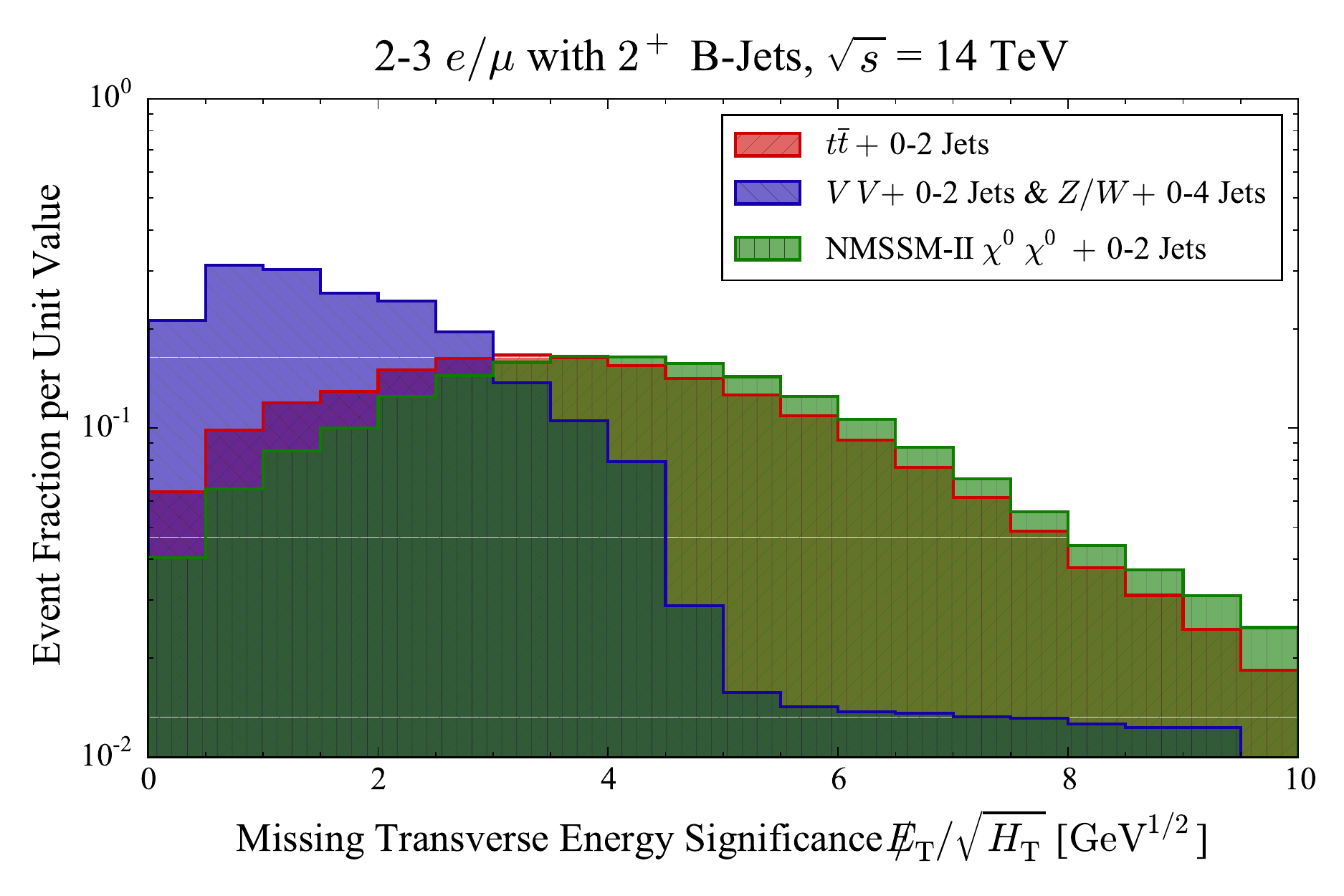}{fig:catII_LEP}
{
Signal and background event shapes are compared for the
final state topology with $2-3$ light leptons and $2^+$ $b$-Jets (category II).
Left: Around 70\% of the
signal features an OS-LF dilepton pair with an invariant mass of
$92\pm5$~GeV, whereas the same holds true for
only approximately 3\% of the $\ttbar$ background component.
Right: The relative (dimensionful) significance of the missing transverse energy
as a numerical ratio of $\met$ to the square-root of the event scale $M_{\rm T}$
(both in GeV) is substantially larger for the signal (as well as for $\ttbar$)
than the vector background components.
}

\PlotPairWide{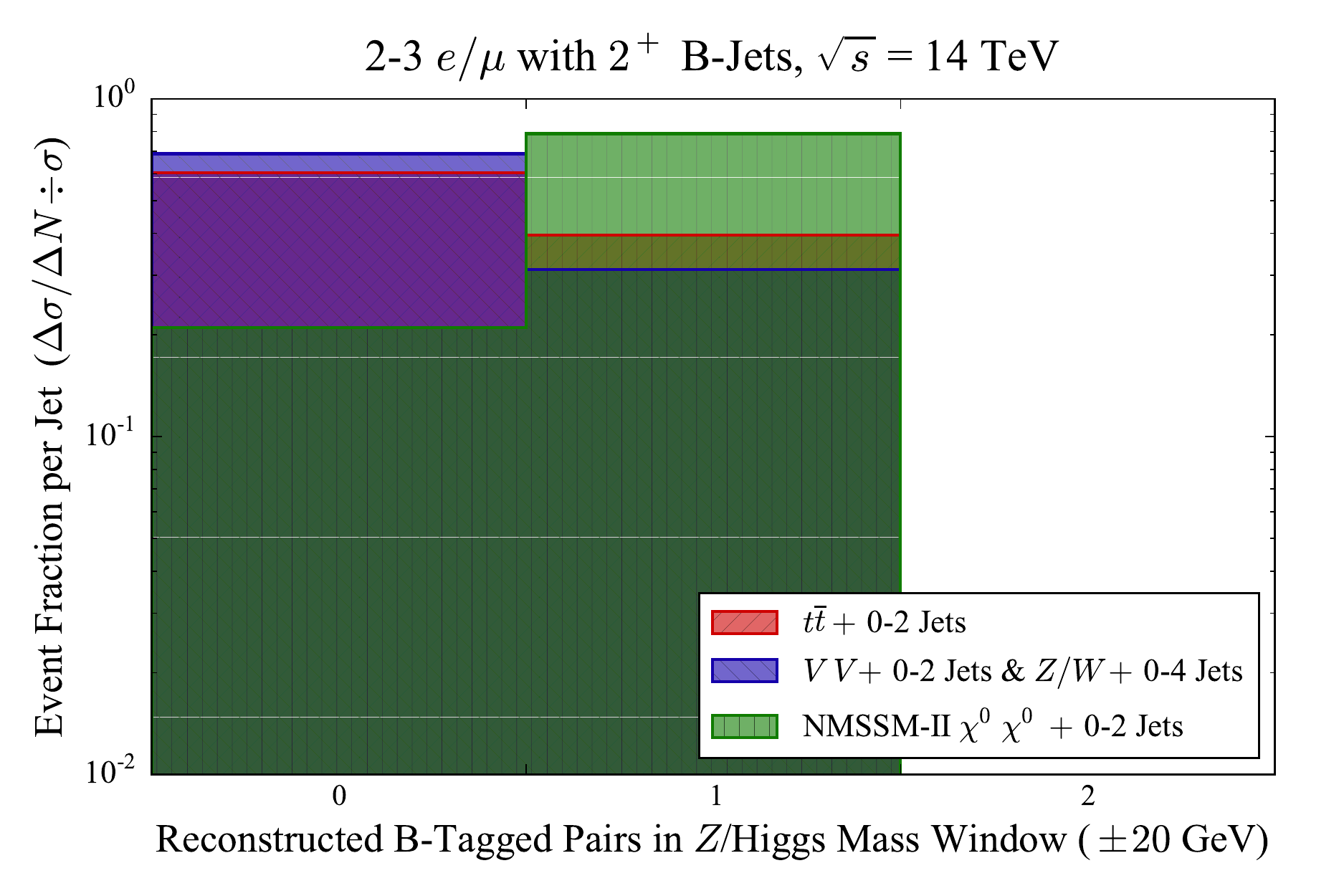}{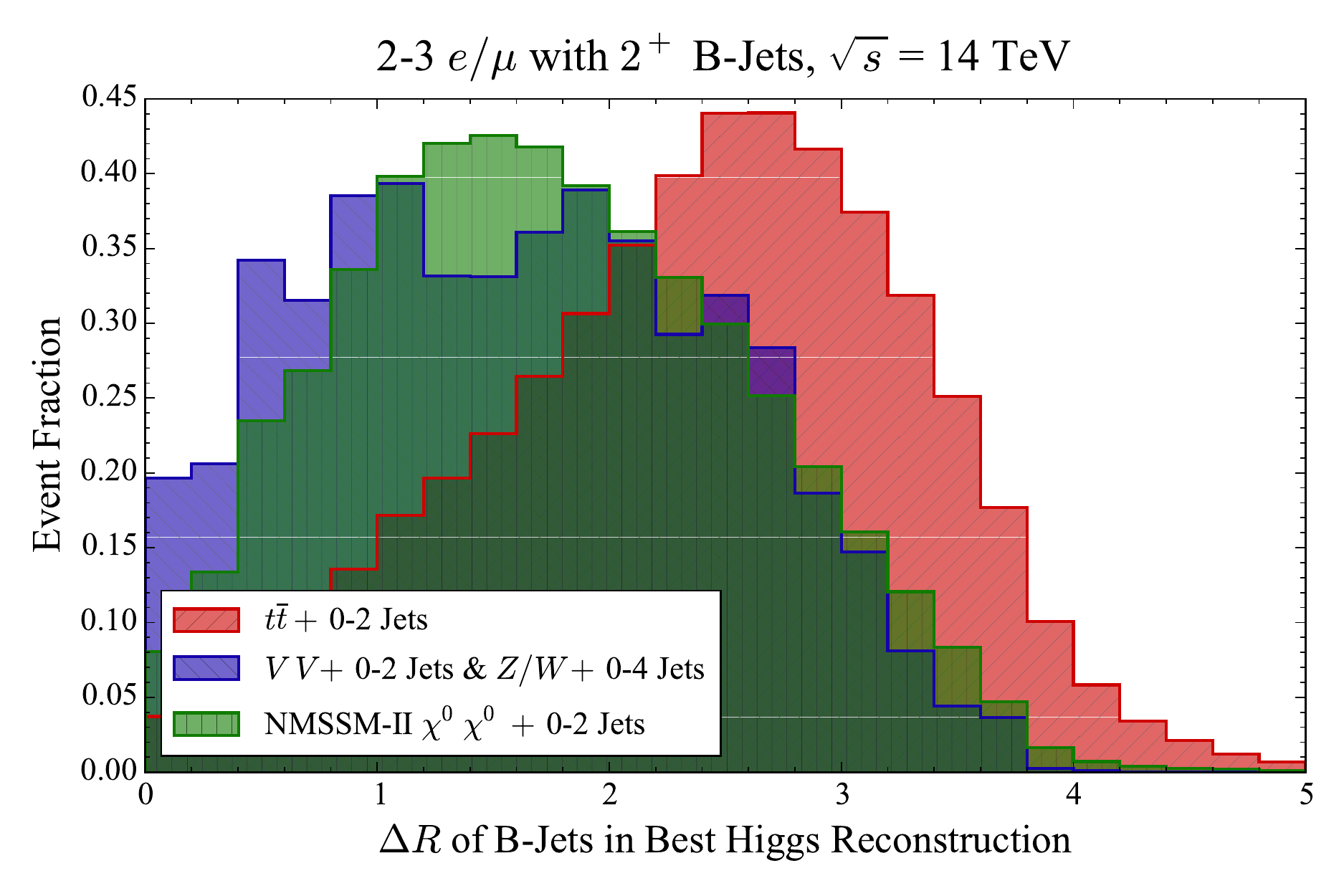}{fig:catII_JET}
{
Signal and background event shapes are compared for the
final state topology with $2-3$ light leptons and $2^+$ $b$-Jets (category II).
Left: About 80\% of the signal features a reconstructed $b$-Jet pair
in the $Z/H$-boson mass window ($92-20$~GeV to $126+20$~GeV),
whereas the same holds true for just 30--40\% of the unified background components.
Right: The angular separation $\Delta R$ of the pair of jets that come closest by
invariant mass to reconstructing a Higgs is systematically smaller for the signal
than the $\ttbar$ background component.
}

\PlotPairWide{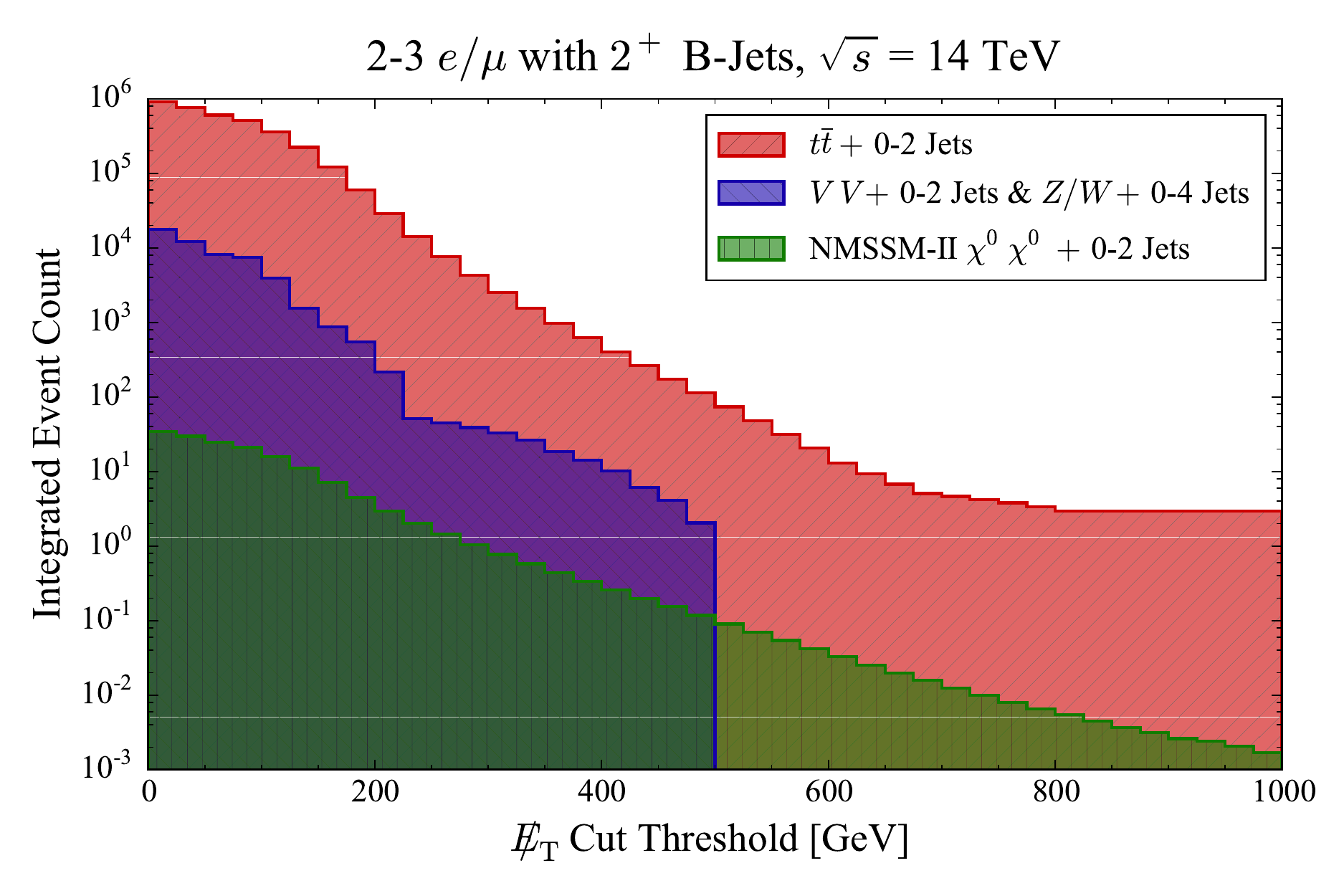}{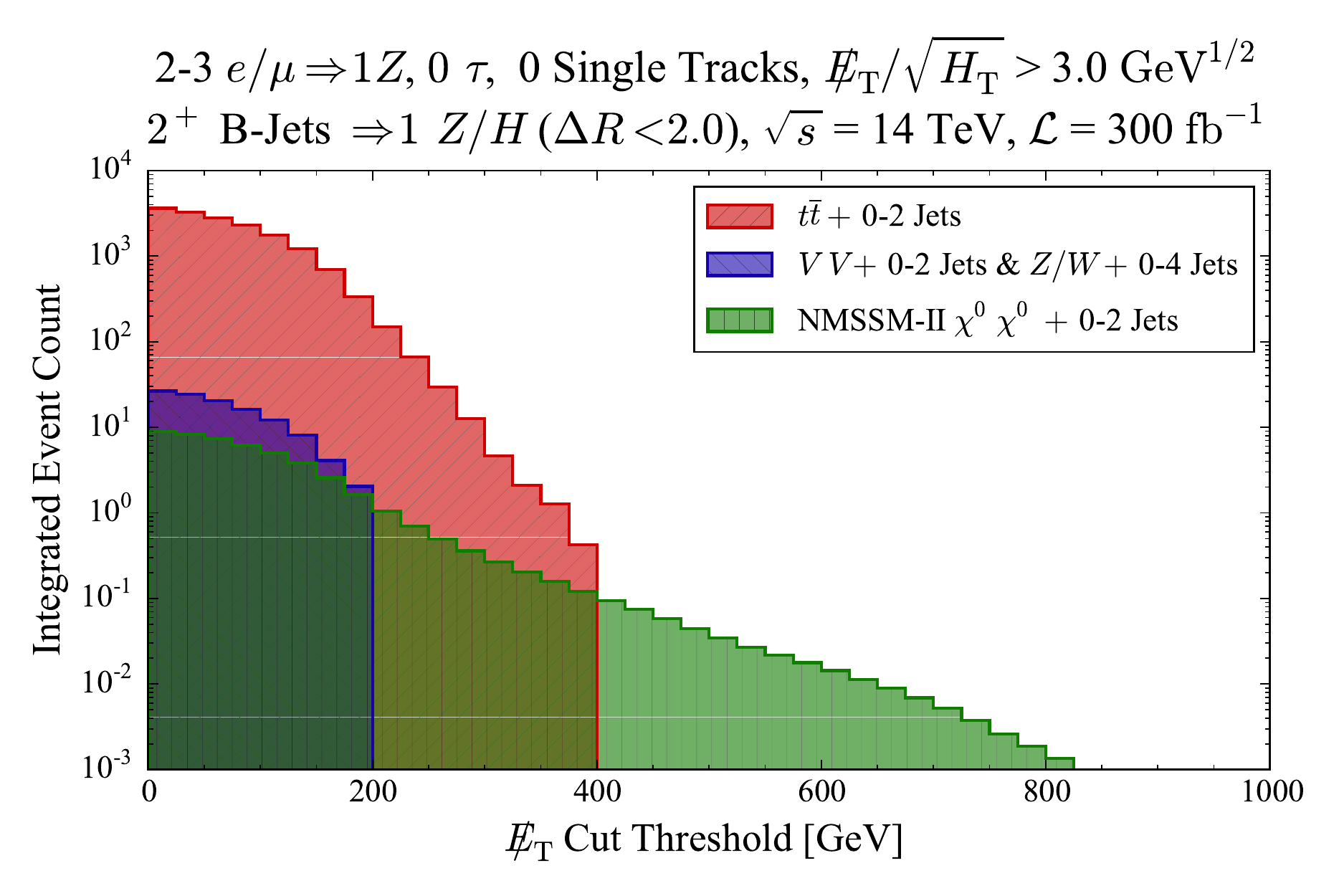}{fig:catII_MET}
{
Signal and background integrated event counts are compared for the
final state topology with $2-3$ light leptons and $2^+$ $b$-Jets (category II) at a luminosity of 300
events per femtobarn as a function of the missing transverse energy $\met$ cut threshold.
Left: The raw event categorization reveals daunting background domination
by $\ttbar+$Jets, with no substantive improvement in the signal to background
ratio at large values of the missing energy.
Right: Enacting the secondary event selections (0 $\tau$, 1 leptonic $Z$, 0 single-track jets,
1 hadronic $Z/H$ with $\Delta R < 2.0$, and $\met/\sqrt{H_{\rm T}} > 4.0$), the signal
to background ratio is improved by two or three magnitude orders (more at larger $\met$
cuts), although it remains apparently intractable at the studied luminosity and signal cross section.
}

\PlotPairWide{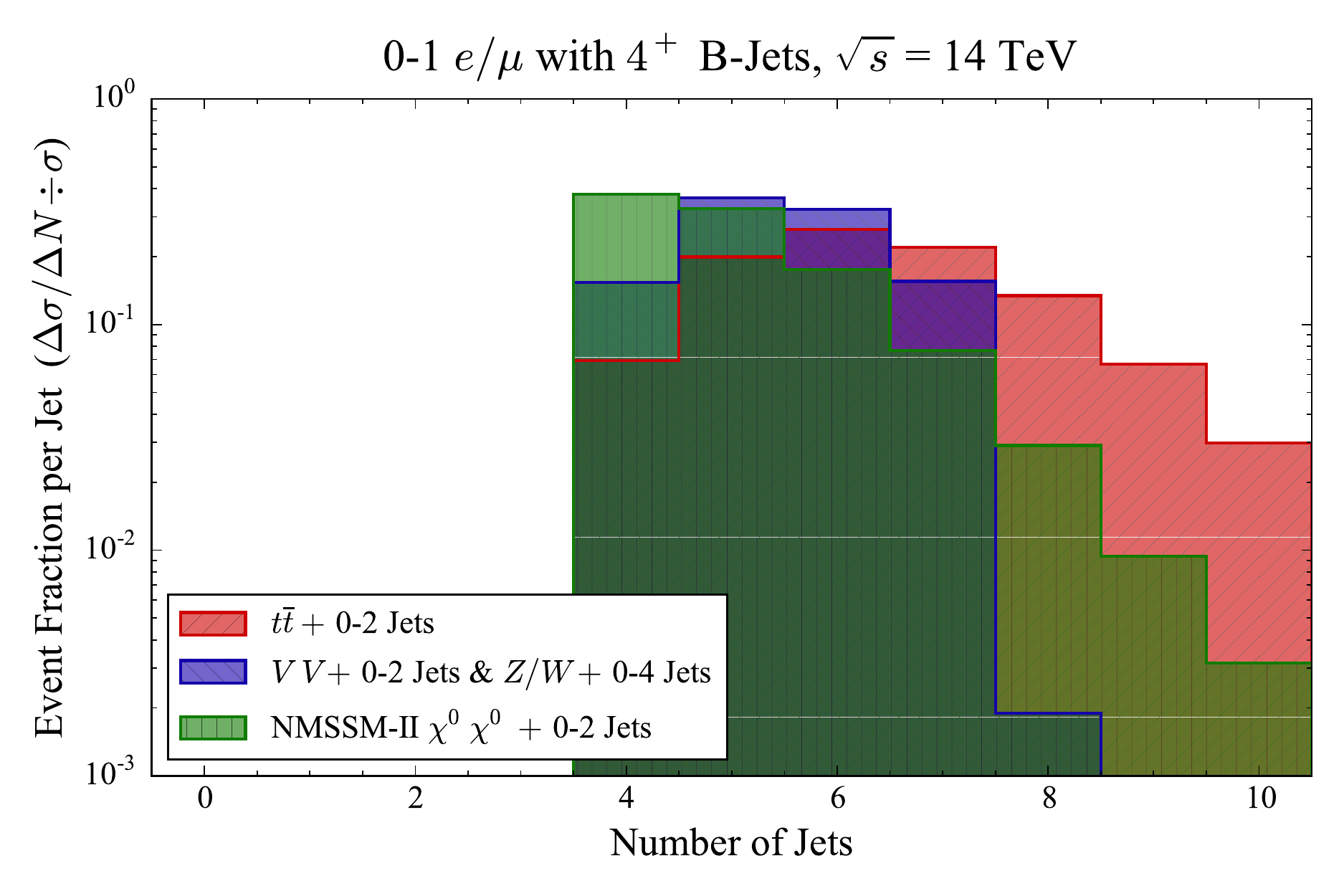}{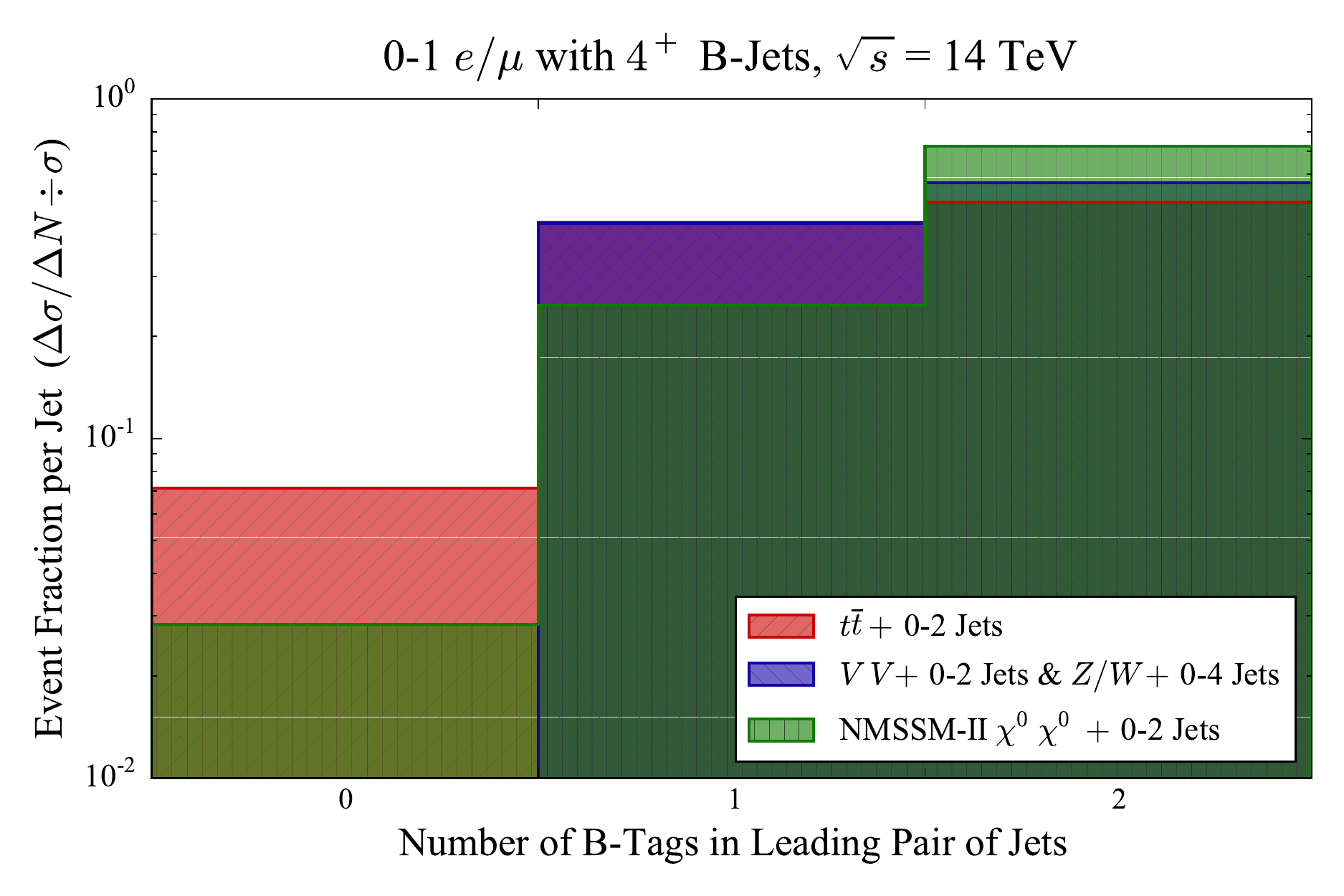}{fig:catIII_JET}
{
Signal and background event shapes are compared for the
final state topology with $0-1$ light leptons and $4^+$ $b$-Jets (category III).
Left: The $\ttbar$ background is generally jettier than the signal, with
a larger fraction of events at six or more jets.
Right: The leading pair of signal jets is somewhat more likely to be $b$-tagged
than the leading pair of jets in the $\ttbar$ background, one or more of
which are likely to be initial state radiation.
}

\PlotPairWide{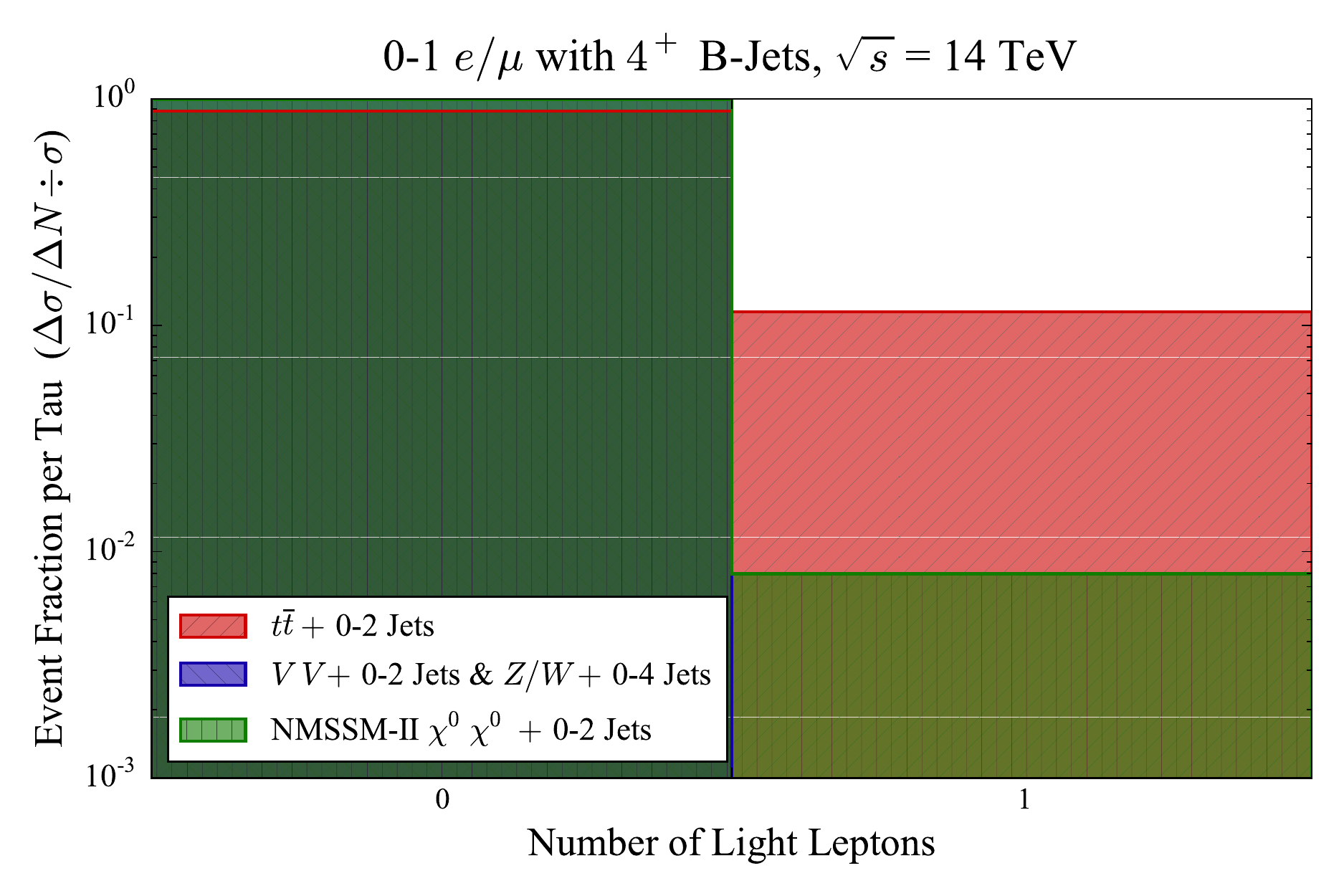}{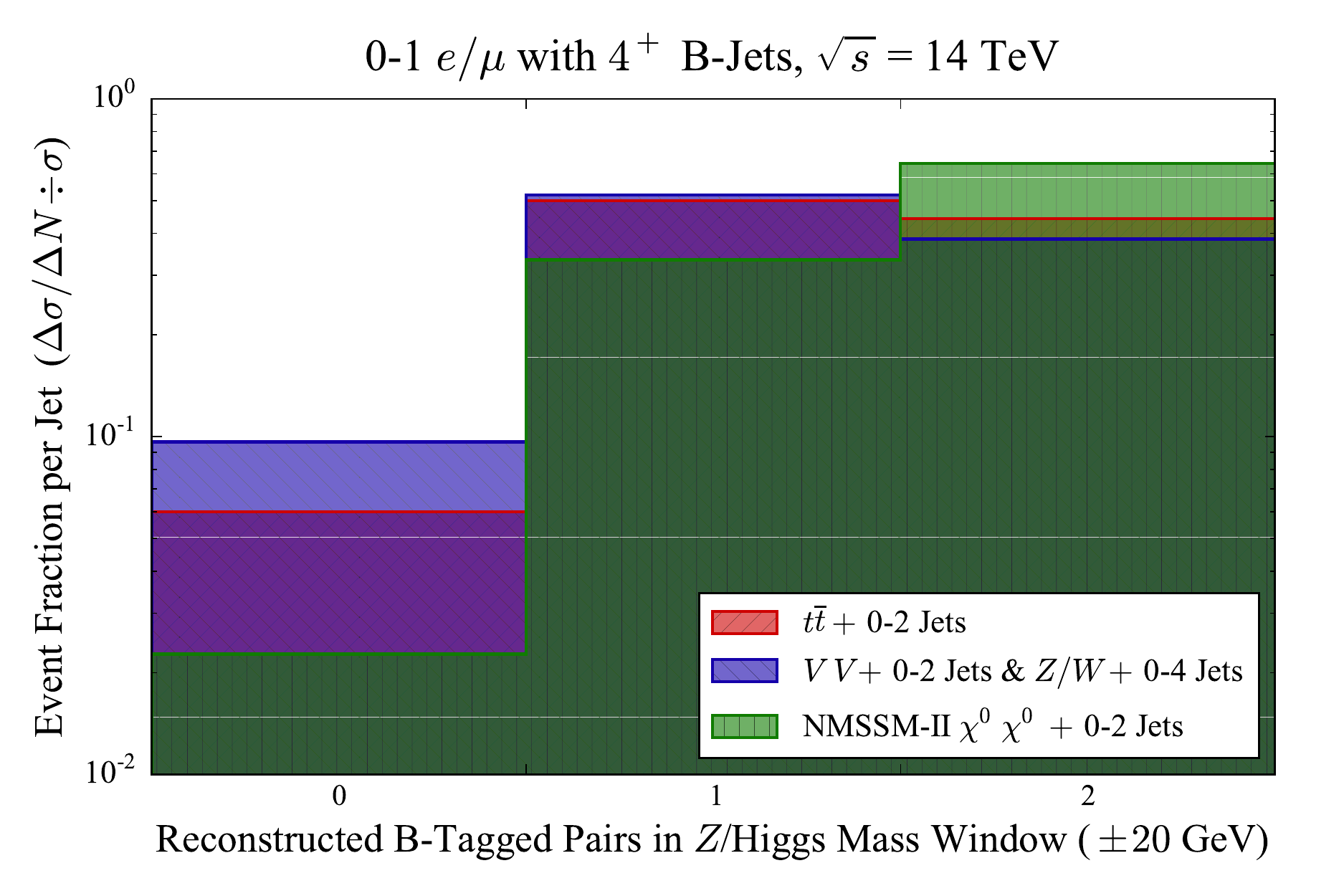}{fig:catIII_LEP}
{
Signal and background event shapes are compared for the
final state topology with $0-1$ light leptons and $4^+$ $b$-Jets (category III).
Left: The $\ttbar$ background is substantially more likely to retain a
single lepton than the signal.
Right: Around 65\% of the signal features two reconstructed $b$-Jet pairs in
the $Z/H$-boson mass window ($92-20$~GeV to $126+20$~GeV),
whereas the same holds true for just around 40-45\% 
of the $\ttbar$ (vector) background components.
}

\PlotPairWide{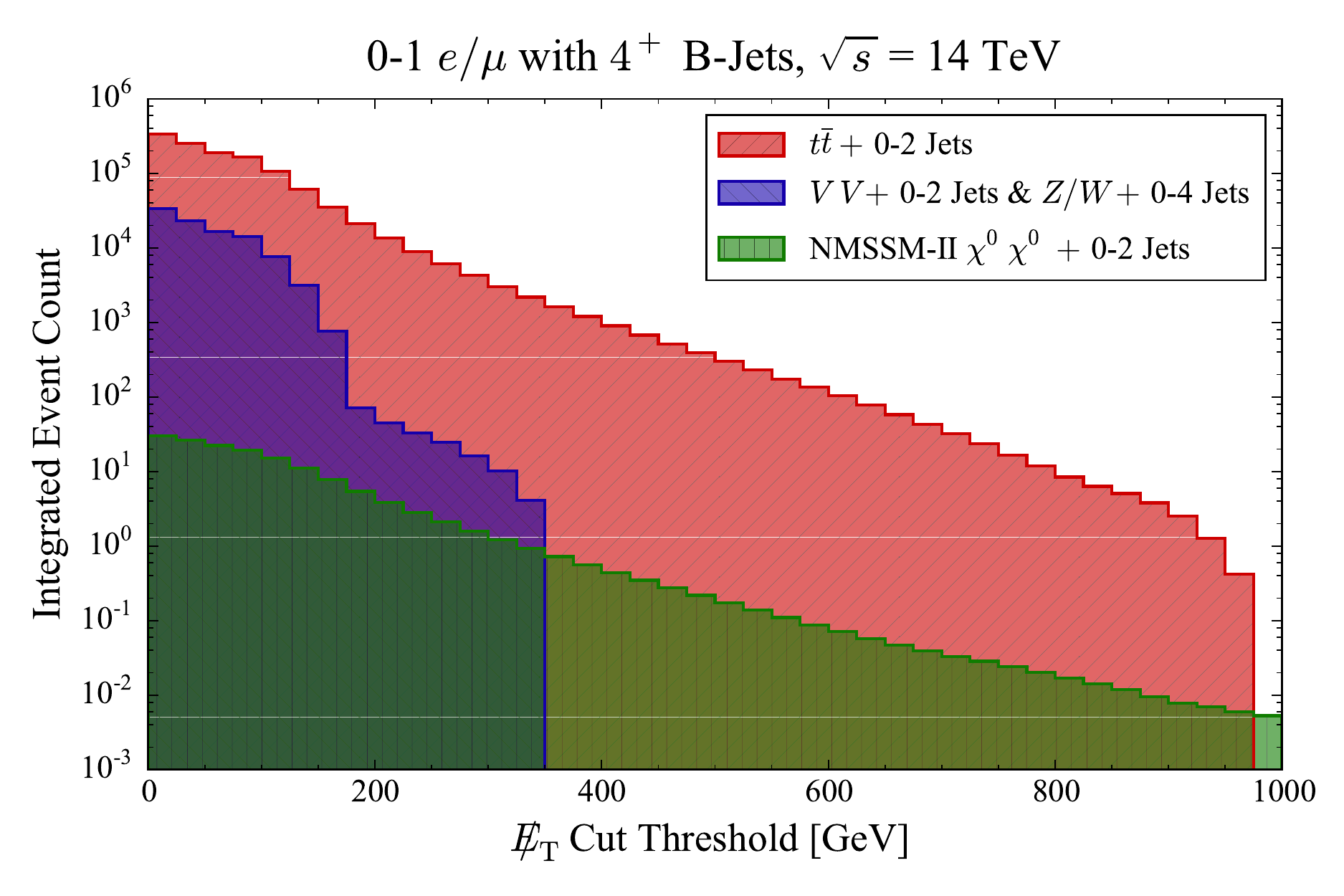}{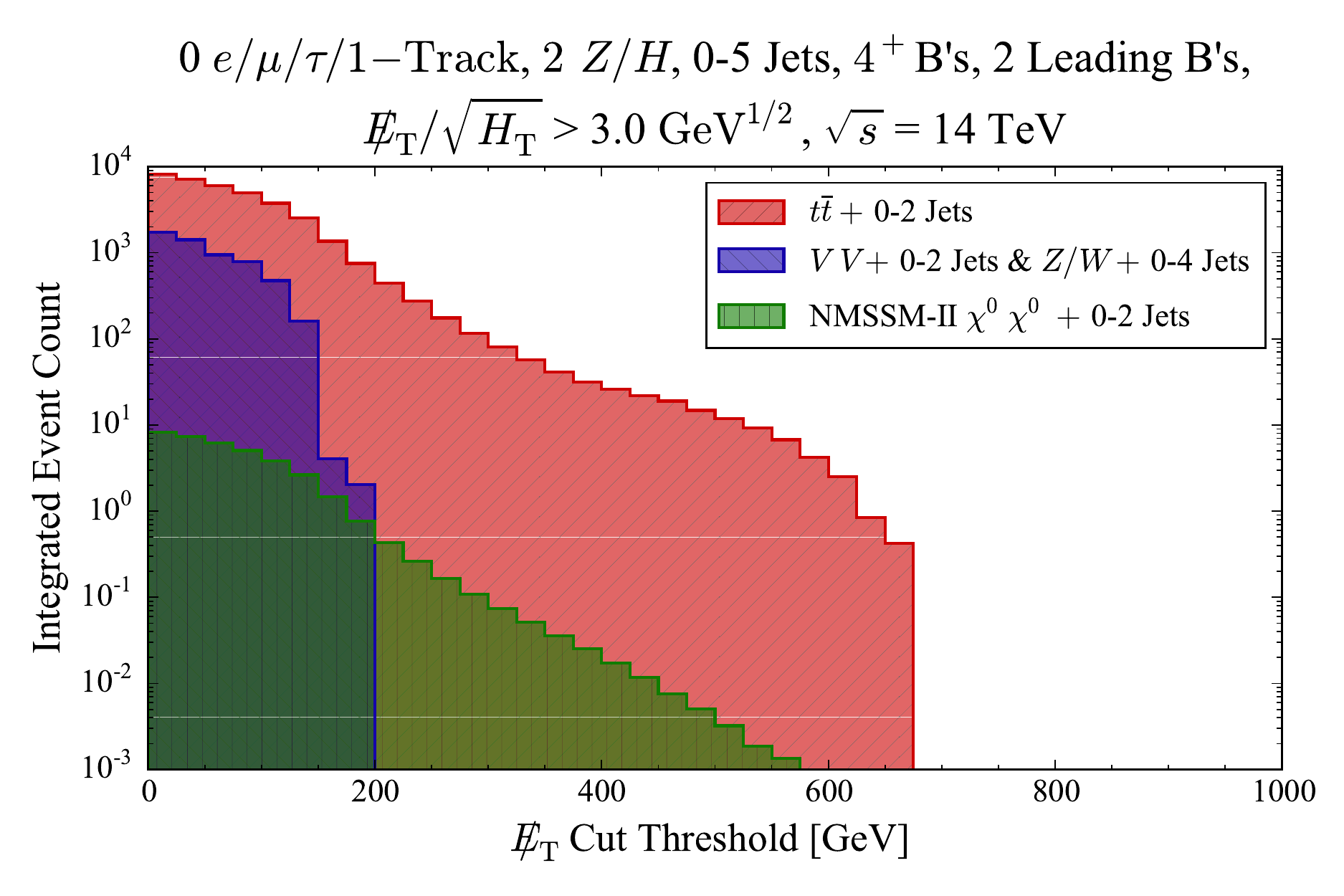}{fig:catIII_MET}
{
Signal and background integrated event counts are compared for the
final state topology with $0-1$ light leptons and $4^+$ $b$-Jets (category II) at a luminosity of 300
events per femtobarn as a function of the missing transverse energy $\met$ cut threshold.
Left: The raw event categorization reveals daunting background domination
by $\ttbar+$Jets, with no substantive improvement in the signal to background
ratio at large values of the missing energy.
Right: Enacting the secondary event selections (0 $e/\mu$, 0 $\tau$, 0-5 total jets,
2 leading $b$-Jets, 0 single-track jets, 2 hadronic $Z/H$, and $\met/\sqrt{H_{\rm T}} > 4.0$),
the signal to background ratio is improved by around a magnitude order,
although it remains apparently intractable at the studied luminosity and signal cross section.
}

\PlotPairWide{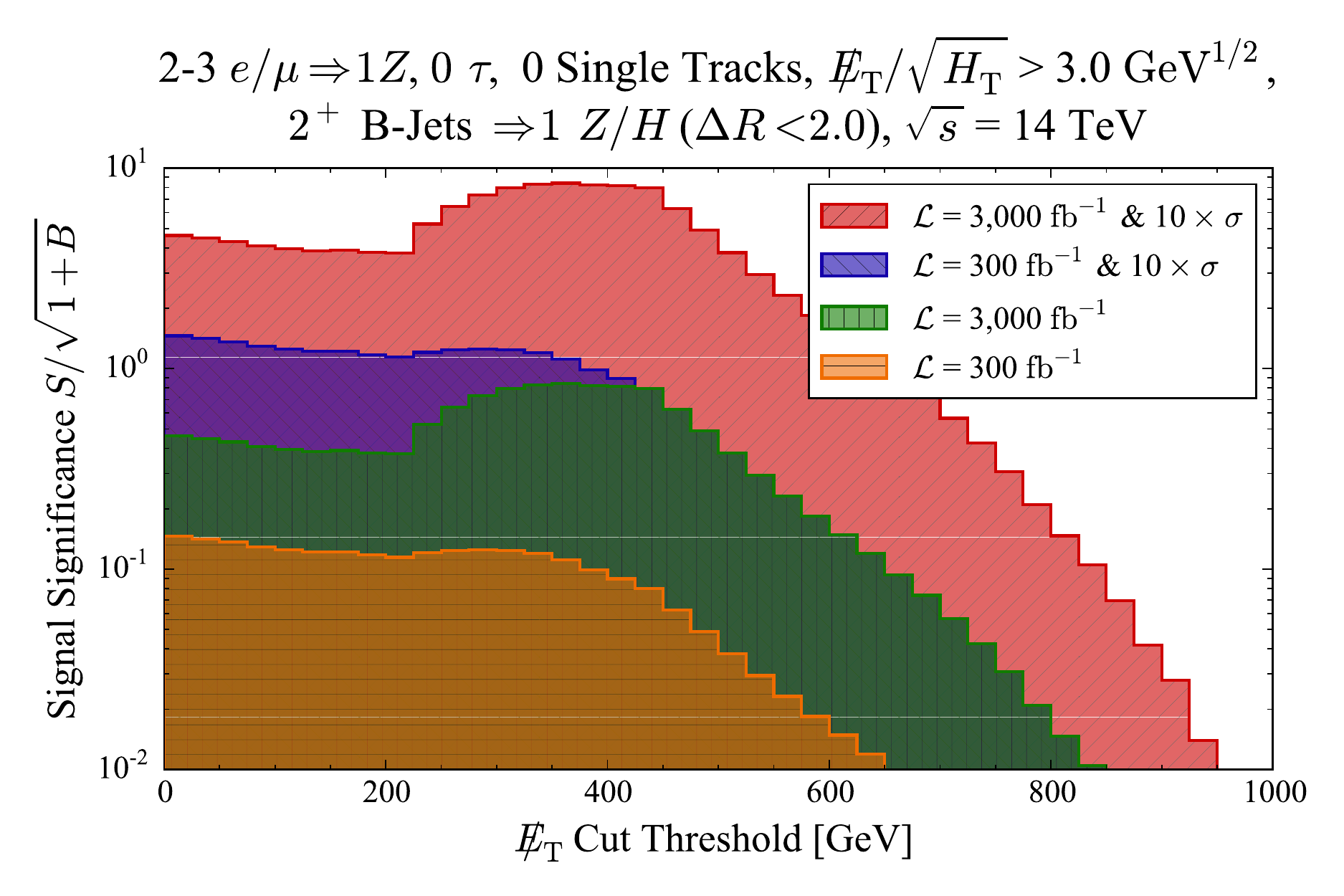}{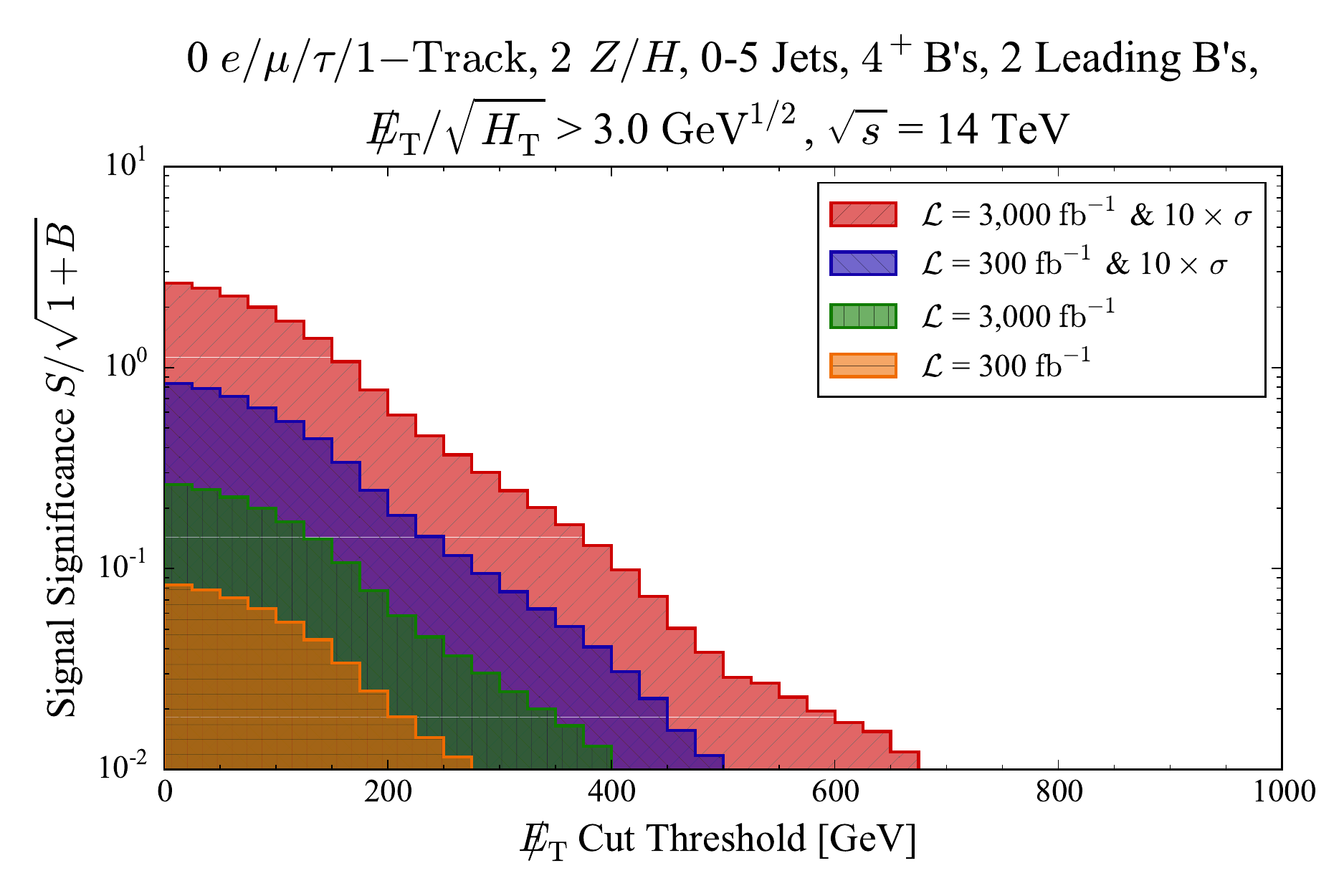}{fig:VAR_OPT}
{
The signal to background significance metric $S/\sqrt{1+B}$ is evaluated
as a function of the missing transverse energy cut threshold for the category II (Left)
and III (Right) final state topologies, applying the optimizations described in the
right-hand panels of Figs.~(\ref{fig:catII_MET},\ref{fig:catIII_MET}). Four contours
are shown, corresponding to luminosities of 300 ${\rm fb}^{-1}$ and 3,000 ${\rm fb}^{-1}$,
for the baseline cross section of the benchmark model and also for a hypothetical spectrum
that is sufficiently more light to engender a one magnitude order increase in the production cross section.
Only by conspiracy of both factors may a significant excess be observed, and then
only for category II. In this former case, a harder cut on $\met > 400$~GeV is suggested if
luminosity and cross section are large enough to support it, in which case background is deeply
contained and the signal is quite visible.  By contrast, in the latter case similarities in the signal and
background missing transverse energy shapes render a substantive $\met$ cut ineffective.
}

The Category II event classification with 2-3 leptons and 2 or more $b$-Jets has the
disadvantage of a final state topology that is readily mimicked by the dual leptonic
decay of $W$ bosons from $\ttbar$ production. With a moderate fake rate for $b$-Jets, the
background is likewise heavily represented in the Category III 0-1 lepton with 4 or more $b$-Jets event topology.
Moreover, the light SUSY electroweak
sector considered for the signal benchmarks does not typically yield a quantity of missing energy
that is sufficiently large to substantially distinguish it from the background. See
Figs.~(\ref{fig:catII_MET},\ref{fig:catIII_MET}), left-hand panels, for a comparison of the raw
event residuals as a function of missing transverse energy cut threshold at a luminosity
of 300~${\rm fb}^{-1}$. The signal is observed to be dwarfed in both cases by around four
magnitude orders. A large variety of kinematic discriminants and specialized discovery variables
have been tested in an effort to identify handles effective for the isolation of signal events.

Considering first category II, it is apparent in the left-hand panel of Figs.~(\ref{fig:catII_LEP}) that the signal is
emphasized by insisting that an available pair of light OS-LF leptons kinematically
reconstruct the mass of a Z-boson (left panel).
The right-hand panel of Figs.~(\ref{fig:catII_LEP}) demonstrates that the 
$\met$ significance variable is again effective at curtailing the vectors plus jets background,
although it is of limited efficacy against the $\ttbar$+Jets background;
we shall select the relatively more modest implementation $\met/\sqrt{H_{\rm T}} > 3.0~{\rm GeV}^{1/2}$
in order to not sacrifice too much signal.
Figs.~(\ref{fig:catII_JET}) show a similar preference for hadronic
($b$-Jet) reconstruction of a particle in the Z/Higgs mass window (left panel), with a narrow constituent
separation in $\Delta R$ (right panel); a cut $\Delta R < 2.0$ will be selected.
As marginal cuts we will opt to also veto single-track jets and hadronic $\tau$'s. 
Figs.~(\ref{fig:catII_MET}) compare the signal and background event residuals before (left) and after (right)
these secondary selections. The background is reduced by more than two magnitude orders, while the
signal is reduced only by a simple factor of 3 or 4. However, the ratio still heavily favors the
$\ttbar+$Jets background component, by 2-3 orders of magnitude (less at higher $\met$ cut thresholds).
The positive response to a cut on missing energy suggests that this could be an effective strategy
if very large luminosities and/or enhanced signal cross-sections were available.

Potential discriminants tested but found to be of limited help in this case include
$M_{\rm T2}$ (the ``s-transverse mass'')~\cite{PAS-SUS-12-002,Lester:1999tx}, 
$M_{\rm T2}^{\rm W}$~\cite{Bai:2012gs},
the jet and dilepton-Z transverse energy balance $\Delta E_{\rm T}$~\cite{Chatrchyan:2012qka},
the razor variables~\cite{Chatrchyan:2011ek,Rogan:2010kb},
the $\alpha_{\rm T}$ ratio~\cite{PAS-SUS-08-005,Randall:2008rw},
the ``biased'' azimuthal difference $\Delta \phi{}^{\rm {\textstyle *}}$~\cite{PAS-SUS-09-001},
the lepton W-projection $L_{\rm P}$~\cite{Chatrchyan:2011ig,PAS-SUS-11-015},
and various transverse thrust and event shape
statistics~\cite{Aad:2012np,Khachatryan:2011dx,Banfi:2004nk,Guchait:2011fb}.

Category III presents similarly in many regards, and faces the same central obstacle that the hadronic
event shape is excessively similar to the background. A similar preference is observed for
a mild cut on the missing energy significance $\met/\sqrt{H_{\rm T}} > 3.0~{\rm GeV}^{1/2}$,
and we again opt to veto on hadronic $\tau$'s and single-track jets.
Distinctions are observed in
Figs.~(\ref{fig:catIII_LEP}), which argues for trimming the total number of jets
to no more than 5 with the leading pair necessarily $b$-tagged, and in Figs.~(\ref{fig:catIII_JET}),
which argues for vetoing light leptons and insisting on two hadronic ($b$-tagged jet)
kinematic reconstructions in the Z/Higgs window. Combinatoric backgrounds
reduce the efficacy of a cut on angular separation for the best mass reconstruction
in this case. Figs.~(\ref{fig:catIII_MET}) compares the signal and background event
residuals before (left) and after (right)
these secondary selections. The background is reduced here by around 1.5 magnitude orders, while the
signal is reduced by a factor close to three. However, the ratio still heavily favors the background,
by about three orders of magnitude, irrespective of a cut on missing transverse energy. 

Figs.~(\ref{fig:VAR_OPT}) evaluate signal to background event significances, using the
metric $S/\sqrt{1+B}$, as a function of the missing transverse energy cut threshold for the category II (Left)
and III (Right) final state topologies, applying the described optimizations
at luminosities of 300 ${\rm fb}^{-1}$ and 3,000 ${\rm fb}^{-1}$,
for the baseline cross section of the benchmark II model and also for a hypothetical spectrum
that is sufficiently more light to engender a one magnitude order increase in the production cross section.
When both scale factors are invoked it appears possible to resolve a significant signal for
the category II final state topology, providing a crucial second data point (in conjunction with the
highly visible first category) for reconstruction of the model. Even granting both factors, the
category III final state topology remains difficult to substantially disentangle from the background.

\section{Discussion and Conclusions}
\label{sect:conclusions}
In this paper we investigated decays from heavy higgsino-like weak-doublets into $Z, h$ bosons and (small) missing
energy. As examples, we considered the MSSM, NMSSM and a singlet-doublet extension of the SM
featuring a DM candidate that is capable of explaining the observed relic density after satisfying direct
detection constraints. The NMSSM is well-motivated by its natural accommodation of the 125 GeV
Higgs and a weak scale value of $\mu$.

Signals from the MSSM, NMSSM and the singlet-doublet extension will be similar, i.e.,
we will expect to find decays into $Z,\,h$ plus missing energy
that cascade into final states with  $4\ell,2\ell 2b$ and $4b$ plus missing energy.
Leptonic products will be dominantly associated with decays of the $Z$, whereas decays of the Higgs will
be associated dominantly with heavy flavor jets.  Establishing two of the prior three final states would
potentially provide a mechanism for quantifying the $Z$ to $h$ ratio, which may in turn
assist in discriminating between specific models exhibiting the described spectral features.
In particular, observation of this ratio will clarify the manner in which the Goldstone Equivalence
Theorem is manifest within and places constraint upon new physics.
The ratio of $Z$ to $h$ production in higgsino-like decays would be somewhat greater than 1
in the case of the MSSM and the singlet-doublet extension, since the net $Z$
rate includes also the contribution of transverse polarizations.
By contrast, in the NMSSM the lighter neutralinos can decay into another state $a$, which can naturally
be close to the light Higgs mass, that masquerades as the Higgs in decays and gives rise
to a $Z/h$ ratio that is smaller than 1 by its.

We explored the visibility of the $4\ell,2\ell 2b$ and $4b$ final states, which are useful
to establishing the $Z/h$ ratio, at the 14 TeV LHC.
If heavy colored particles are to be probed at the LHC, then the lighter MSSM and NMSSM neutralinos and charginos
(or their new fermion counterparts) will likewise be within reach for direct production.
However, the reach for these neutralinos and new fermions is not very high, and
existing bounds vanish rapidly for scenarios with a massive lightest neutralino/fermion.
We selected a representative NMSSM benchmark within this class of models for detailed collider simulation,
with higgsino next-to-lightest neutralinos around 270 GeV, a singlino lightest supersymmetric particle around 140 GeV,
and a light pseudoscalar around 160 GeV.  Leading backgrounds were also simulated, and various
event selection scenarios were tested in an effort to optimize the targeted signals.
The inclusion of 1-2 initial state jets can be helpful in providing some additional
boost to the visible system, although low signal rates, lightness of the invisible final state,
and narrowness of the mass hierarchy were found to limit the efficacy of hard cuts on missing energy.
The four-lepton signal region is substantially visible, with just 
300 fb$^{-1}$ of integrated luminosity proving almost sufficient for a
5$\sigma$ level discovery of the benchmark model.
For the same masses, the $2\ell2b$ and $4b$ final states contend with standard model
backgrounds that prove difficult to reduce.  Discovery is possible in the $2\ell2b$ topology
if the benchmark cross section is elevated by a factor of around ten, in conjunction with
an elevation of the luminosity to the order of 3000 fb$^{-1}$.
Visibility of the $4b$ topology would seem to require a new collider environment,
with substantially upgraded luminosity and/or center-of-mass energy.

It appears that it will be possible for the LHC to establish (in the $4\ell$ channel) the studied class of models,
where only lighter weak fermions exist around the electroweak scale, where the mass gap separating
the dark matter candidate from the next to lightest states is not much larger than $\sim 125$.
For the lightest of these scenarios, and utilizing high luminosities, it seems further
possible that the LHC will be able to also confront the $2\ell2b$ channel, allowing
for direct discrimination of the $Z$ to $h$ ratio in decays of a higgsino-like state.
These observations would function as a probe of the manner in which the new physics
manifests the Goldstone equivalence theorem, and would provide the opportunity to
distinguish between the NMSSM and models such as the MSSM or the singlet-doublet extension of the SM.

\section{Acknowledgments}

We would like to thank Teruki Kamon, Nikolay Kolev and Keith Ulmer
for helpful discussions.
BD acknowledges support from DOE grant no. DE-FG02-13ER42020.
YG acknowledges support from the Mitchell Institute for Fundamental Physics and Astronomy.
DS acknowledges support from DOE grant no. DE–FG02–92ER40701, and the Gordon
and Betty Moore Foundation, through grant no. 776 to the Caltech Moore
Center for Theoretical Cosmology and Physics.
JWW acknowledges support from the SHSU Enhancement Research Grant program,
NSF grant No. PHY-1521105, the SHSU Department of Physics,
and the Mitchell Institute for Fundamental Physics and Astronomy.

\bibliography{bibliography}

\begin{card}[htp]
\centering
\ovalbox{%
\begin{minipage}{0.88\linewidth}
\vspace{3pt}
{\scriptsize
\begin{Verbatim}[numbers=left,numbersep=6pt]
******** cut_card.dat v3.15 ***
* Classify Objects with No Cuts
*** Object Reconstruction ****
        # ALL Jets
OBJ_JET_000 = PTM:30, PRM:[0.0,5.0], CUT:0
        # LEAD Jet
OBJ_JET_001 = SRC:+000, PRM:[0.0,2.5],
        CUT:[1,UNDEF,-1], OUT:PTM_001, ANY:0
        # SECOND Jet
OBJ_JET_002 = SRC:[+000,-001], PRM:[0.0,2.5],
        CUT:[1,UNDEF,-1], OUT:PTM_002, ANY:0
        # B-Tagged Jets
OBJ_JET_003 = SRC:+000, PRM:[0.0,2.5], HFT:0.5, CUT:0
        # Non-B Jets
OBJ_JET_004 = SRC:[+000,-003], PRM:[0.0,2.5], CUT:0
        # B-TAGS in Jets 1,2
OBJ_JET_005 = SRC:[+001,+002], HFT:0.5, CUT:0
        # Non-B Sub-Leading Jets
OBJ_JET_006 = SRC:[+000,-001,-002,-003],
        PRM:[0.0,2.5], CUT:0
        # 1 B-Tags in Z/Higgs Window
OBJ_JET_007 = SRC:+003, EFF:[WIN,92,20,126,20,1], CUT:0
        # 2 B-Tags in Z/Higgs Window
OBJ_JET_008 = SRC:+003, EFF:[WIN,92,20,126,20,2], CUT:0
        # 2 B-Tags in Higgs Window
OBJ_JET_009 = SRC:+003, EFF:[WIN,126,20,2], CUT:0
        # Single Track Jets
OBJ_JET_010 = SRC:+000, TRK:[1,1], CUT:0
        # Leading or B-Tagged Jets (No Output)
OBJ_JET_011 = SRC:[+001,+002,+003]
        # Nearest B-Tag Object Pair to Higgs Window
OBJ_JET_012 = SRC:+003, EFF:[OIM,126,UNDEF,-1]
        # Further B-Tag Object Pair from Higgs Window
OBJ_JET_013 = SRC:[+003,-012], EFF:[OIM,126,UNDEF,-1]
        # ALL Leptons
OBJ_LEP_000 = PTM:10, PRM:[0.0,2.5]
        # Light Soft Leptons
OBJ_LEP_001 = SRC:+000, EMT:-3, SDR:[0.3,UNDEF,1], CUT:0
        # Soft Taus
OBJ_LEP_002 = SRC:+000, EMT:+3, CUT:0
        # DiLepton Pairs in Z Window
OBJ_DIL_001 = LEP:001, DLS:-1, DLF:1, WIN:[92,5], CUT:0
OBJ_DIL_002 = LEP:001, DLS:-1, DLF:1, WIN:[92,10], CUT:0
****** Event Selection *******
        # Full Event Missing Transverse Energy
EVT_MET_000 = OUT:1
        # MET-Jet Delta Phi (Leading+B-Tags)
EVT_MDP_001 = MET:000, JET:011, OUT:1
        # MET Significance MET / sqrt( HT )
EVT_RHR_001 = NUM:000, DEN:000, OUT:1
        # Invariant Mass of Nearest Higgs Window Pair
EVT_OIM_001 = JET:012, OUT:1
        # Invariant Mass of Further Higgs Window Pair
EVT_OIM_002 = JET:013, OUT:1
        # Delta-R Separation of Nearest Higgs Window Pair
EVT_ODR_001 = JET:012, OUT:1
        # Delta-R Separation of Further Higgs Window Pair
EVT_ODR_002 = JET:013, OUT:1
****** Event Filtering *******
        # Category I: 4 Leptons, 0+ B-Jets
CUT_ESC_001 = KEY:LEP_001, CUT:4
CUT_ESC_002 = KEY:JET_003, CUT:0
CUT_CHN_001 = ESC:[+001,+002]
        # Category II: 2-3 Leptons, 2+ B-Jets
CUT_ESC_003 = KEY:LEP_001, CUT:[2,3]
CUT_ESC_004 = KEY:JET_003, CUT:2
CUT_CHN_002 = ESC:[+003,+004]
        # Category III: 0-1 Leptons, 4+ B-Jets
CUT_ESC_005 = KEY:LEP_001, CUT:[0,1]
CUT_ESC_006 = KEY:JET_003, CUT:4
CUT_CHN_003 = ESC:[+005,+006]
******************************
\end{Verbatim}
} \vspace{-3pt} \end{minipage}}
\caption{{\sc AEACuS} instruction card for computation of relevant event statistics.
Pre-filtering into event topology categories I-III is performed in the final lines.}
\label{CARD:GENERAL}
\end{card}

\appendix

\section{Doublet decay in the SDF model}
\label{app:Zh_in_SDF}

Assuming $M_S < M_D$, and that the lightest $\tilde{\chi}^0_1$ is mostly the singlet $S$, the heavier $\tilde{\chi}^0_2,\tilde{\chi}^0_3$
are mixtures of the doublets $D_1,D_2$, and the squared matrix elements of their decay processes are
\bea
\scriptsize \label{eq:Msq_Z}&{\cal M}^2_{\chi^0_i\rightarrow\chi^0_jZ}& =
\frac{y^2}{2}\left[ (M_{\chi^0_i}+M_{\chi^0_j})^2-M_Z^2\right] \\
&\times&\hspace{-22pt}\left[(N_{i2}N_{j1}+N_{i1}N_{j2}) c_\theta+ (N_{i3}N_{j1}+N_{i1}N_{j3})s_\theta\right]^2 \nn \\
&+&\hspace{-22pt}\frac{g^2+g'^2}{2}(N_{i2}N_{j2}-N_{i3}N_{j3})^2\nn \\
&\times&\hspace{-22pt}\left( M_{\chi_i^0}^2+M_{\chi_j^0}^2+4M_{\chi_i^0}M_{\chi_j^0}-M^2_Z \right)\nn \\
\label{eq:Msq_h}
&{\cal M}^2_{\chi^0_i\rightarrow\chi^0_jh}& = \frac{y^2}{2} \left[ (M_{\chi^0_i}+M_{\chi^0_j})^2-M_h^2\right] \\
&\times&\hspace{-22pt}\left[(N_{i2}N_{j1}+N_{i1}N_{j2})c_\theta-(N_{i3}N_{j1} - N_{i1}N_{j3})s_\theta\right]^2 \!,\normalsize \nn
\eea

where $s_\theta, c_\theta$ are short for $\sin\theta,\cos\theta$. $N_{ij}$ are the elements of the mixing matrix that
diagonalizes the mass matrix in Eq.~\ref{eq:mmat}. $\chi^0_{2,3}$ either decay into the singlet
component of $\chi^0_1$ via the $ySHD$ terms, or into the small doublet component in $\chi^0_1$ via the gauge couplings. 

Note that without the  $ySHD$ term, i.e. in the limit $y\rightarrow 0$, the singlet would altogether decouple from the
doublet, $\chi^0_1$ would have no mixing into $D_1,D_2$, implying $N_{12}=N_{13}=0$, and the prior decays
would become forbidden. Turing on $y>0$, with $y$ still smaller than unity, the doublet mixings in $\chi^0_1$
grow linearly with $y$, implying $N_{12},N_{13} \propto y$, giving a $y^2$ dependence also in the second
term in Eq.~\ref{eq:Msq_Z}. Thus, decay widths into both $Z,h$ grow as $y^2$ when $y\ll 1$, and thereby maintain
a comparable size. At large $y$, $y\cdot vev\sim M_S, M_D$ in the mass terms, which cause the $D_1,D_2$
mixings to become more complicated.  Still, this property qualitatively holds, as is observed for the
benchmark Point III in Table~\ref{tab:benchmark3}.
\vspace{2cm}

\section{{\sc AEACuS} Event Selection Card}
\label{app:card}

Selection cuts and computation of collider observables
have been implemented within {\sc AEACuS~3.15}~\cite{Walker:2012vf,aeacus}
using the instructions in Card~\ref{CARD:GENERAL}. 

\end{document}